\documentclass[aps,pre,onecolumn,floatfix]{revtex4}

\usepackage{graphicx}
\usepackage{amssymb,amsfonts,amsmath}
\usepackage{mathtools}
\usepackage{epsf}
\usepackage{subfigure}
\usepackage{epstopdf}

\newcommand{\be}{\begin{equation}}
\newcommand{\ee}{\end{equation}}
\newcommand{\bea}{\begin{eqnarray}}
\newcommand{\eea}{\end{eqnarray}}

\newcommand{\dd}{{\mathrm d}}

\begin{document}

\title{\bf Kinetics of multistate DNA polymerases}

\author{Pierre Gaspard}
\affiliation{Center for Nonlinear Phenomena and Complex Systems,\\
Universit\'e Libre de Bruxelles (ULB), Code Postal 231, Campus Plaine,
B-1050 Brussels, Belgium}

\begin{abstract}
In the present paper, we apply the iterative mathematical method previously developed for the kinetics of template-directed multistate copolymerization to the kinetics of DNA replication by polymerases having multiple structural states.  In particular, we study a two-state kinetic model for the T7 DNA polymerase.  We obtain the mean velocity for the growth of the copy along the template, the error probability of DNA replication by the polymerase, and the local probabilities of base-pair formation along the template sequence.  Furthermore, we show that the iterative method is more than a million times faster than usual numerical simulation methods.  Results are also obtained in the approximation of homogenization of template heterogeneities.
\end{abstract}

\noindent 
\vskip 0.5 cm

\maketitle

\section{Introduction}

In the companion paper~\cite{paperI}, we discussed the kinetics of template-directed multistate copolymerization processes, using a mathematical method based on matrix iterations, which gives the exact asymptotic solution of the kinetic equations of copolymerization.  In the present paper, we apply this mathematical method to study the kinetics of multistate DNA polymerases like the one of bacteriophage T7  \cite{TJ06,J10,DJ20,DJ21,DKJ22}.

In the study of DNA replication, DNA polymerases are considered as enzymes obeying a Michaelis-Menten kinetics with fast binding and unbinding of nucleotides, followed by a slower polymerization step \cite{MM13,JG11,S81,L82}.  A basic assumption, which was used in early work, consists in neglecting the effects of template heterogeneities and approximating the kinetics by homogenization of the process into the formation of correct and incorrect base pairs \cite{PWJ91,WPJ91,J93}.  Nowadays, the experimental methods of biochemistry are able to measure the kinetic parameters for each one of the $4\times 4$ possible base pairs, which can be formed during replication \cite{AWT97,FS04,LJ06,ZBNS09,BBT12,SG15}.  These measurements were carried out for several DNA polymerases, showing that they may often be assumed to have a single structural state to understand their kinetics.  However, for high-fidelity DNA polymerases, experimental observations revealed large changes in conformation between an open structural state in the absence of nucleotide and a closed state after binding nucleotide \cite{J10}.  These conformational changes are thought to play an important role for high replication fidelity \cite{DJ20}.

In this regard, the theory of DNA replication should be extended from single-state to multistate processes.  As shown in the companion paper~\cite{paperI}, this extension can be carried out in great generality, providing the exact mathematical solution of the kinetic equations in the long-time limit, using an iterative method, which is much faster than usual methods of numerical simulation.  This iterative method is here applied to a two-state model of the T7 DNA polymerase, where the replication process only involves two species of nucleotide, namely, adenosine~A and thymine~T.  For this model, we can use the experimental data of Refs.~\cite{DJ20,DJ21,DKJ22} to infer values for the kinetic parameters.  The results of the iterative method are compared to those of numerical simulation with Gillespie's algorithm \cite{G76,G77}.  This comparison confirms that the two methods give the exact solution of the kinetic equations, but the theoretical iterative method is computationally much faster than the method of numerical simulation.  Furthermore, relationships can be established with the mathematical formulas used in biochemistry \cite{J10,DKJ22}.

The plan of this paper is the following.  Section~\ref{sec:math} presents the two-state kinetic model of DNA polymerases.  The kinetic equations of this model are obtained under the conditions of Michaelis-Menten enzyme kinetics.  In Section~\ref{sec:theory}, the kinetic equations are exactly solved using the iterated matrix function system (IMFS) method developed in the companion paper~\cite{paperI}.  Remarkably, the matrix iterations reduce to scalar iterations for the kinetic process of the two-state model we consider, leading to a standard iterated function system \cite{BD85}, as for single-state kinetics \cite{G16PRL,G17PRE} but with significant differences.  The numerical results are given in Section~\ref{sec:results}, comparing the theoretical iterative method to the numerical method of Gillespie's algorithm.  Section~\ref{sec:homog} is devoted to the approximation of homogenization for the process of DNA replication.  Starting from the full model of copolymerization along a heterogeneous template, homogenization leads to a simplified model where only correct and incorrect pairs are considered.  In this way, we can obtain approximate but analytical expressions, which are very useful for quantitative comparisons.  Conclusion and perspectives are given in Section~\ref{sec:conclusion}.  Appendix~\ref{AppA} gives the deduction of the kinetic equations using the Michaelis-Menten conditions.  The sources of the values adopted for the kinetic parameters of the model are presented in Appendix~\ref{AppB}.  The regime of sublinear growth in time is discussed in Appendix~\ref{App:Sublin}.

\section{Model and equations for the kinetics of DNA polymerases}
\label{sec:math}

\subsection{The kinetic model}

We study a model of high-fidelity DNA polymerases having two structural states, an open state E and a closed state F, as experimentally observed \cite{DJ20,DJ21,DKJ22}.  For such polymerases, DNA replication proceeds as follows:
\be
{\rm E}\cdot{\rm DNA}_l \ + \ {\rm dNTP} 
  \quad {\overset{1}{\rightleftharpoons}}
  \quad
  {\rm E}\cdot{\rm DNA}_l\cdot{\rm dNTP} 
  \quad {\overset{2}{\rightleftharpoons}}
  \quad
  {\rm F}\cdot{\rm DNA}_l\cdot{\rm dNTP} 
  \quad {\overset{3}{\rightleftharpoons}}
  \quad
{\rm E}\cdot{\rm DNA}_{l+1} \ + \ {\rm PP}_{\rm i} \; ,
\label{pol-react}
\ee
where $l$ is the length of the DNA strand, dNTP is either a correct or an incorrect deoxyribonucleoside triphosphate, and PP$_{\rm i}$ is the inorganic pyrophosphate that is released after polymerization, i.e., the formation of the phosphodiester bond between the new nucleotide unit and the already grown strand.  As for other enzymes obeying Michaelis-Menten kinetics \cite{MM13,JG11,S81,L82}, the reaction step~1 is at quasi-equilibrium because of rapid binding and unbinding of a deoxyribonucleoside triphosphate with hydrogen bonds to a template unit.  The step~2 is the conformational change of the polymerase between its structural states~E and~F.  The step~3 is the polymerization reaction with the release of pyrophosphate and the translocation after polymerase opening.  The reversible reactions~(\ref{pol-react}) tend to proceed from left to right if the deoxyribonucleoside triphosphate concentrations are large enough with respect to the pyrophosphate concentration.  The processivity of the polymerase is assumed to be infinite and to run on an arbitrarily long template sequence $n_1n_2\cdots n_l n_{l+1}\cdots$ [which is not shown in the scheme~(\ref{pol-react})], while synthesizing a copy ${\rm DNA}_l=m_1m_2\cdots m_l$.  We consider the simple case where the template and the copy are composed of only two types of nucleotides, adenosine~A and thymine~T, so that their units are $m_l,n_l\in\{{\rm A},{\rm T}\}$.  Furthermore, the solution surrounding the polymerase only contains the deoxyribonucleoside triphosphate dATP and dTTP at given concentrations, which are the control parameters of the process.  The template sequence is taken as Bernoulli chain with the probabilities $0\le \nu({\rm A})\le 1$ and $\nu({\rm T})=1-\nu({\rm A})$.

\begin{figure}[h]
\includegraphics[width=8cm]{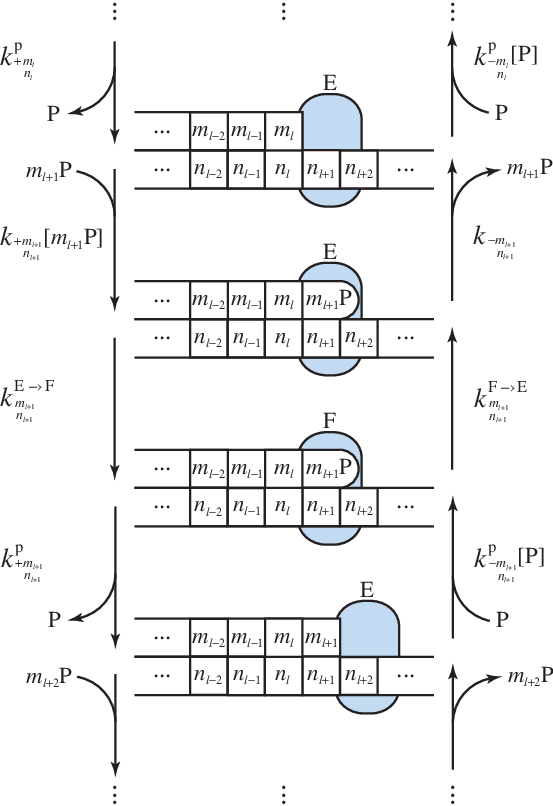}
\caption{Representation of the kinetic scheme for the model of two-state DNA polymerases.  E denotes the open state of the polymerase, F its closed state, $\{m_j\}$ the nucleotide units composing the copy, $\{n_j\}$ those of the template, $\{m_j{\rm P}\}$ the molecules of deoxyribonucleoside triphosphate, and P the molecules of inorganic pyrophosphate.  The rates of the reactions are explained in the text.}
\label{fig1}
\end{figure}

The different steps of the template-directed copolymerization process are shown in detail in Fig.~\ref{fig1}, where $m_l\in\{{\rm A},{\rm T}\}$ are the copy units, $n_l\in\{{\rm A},{\rm T}\}$ those of the template, $m_l{\rm P}\in\{{\rm dATP},{\rm dTTP}\}$ the molecules of deoxyribonucleoside triphosphate with the concentrations $[m_l{\rm P}]$ in the surrounding solution, and P the pyrophosphate molecules with the concentration $[{\rm P}]$.  In Fig.~\ref{fig1}, we denote the rate constants of polymerization and depolymerization of the copy unit $m_l$ paired with the template unit $n_l$ by $k^{\rm p}_{\pm m_{l}\atop \ \, n_{l}}$; the rate constants of nucleotide binding and unbinding to form the pair $m_l\!:\!n_l$ by $k_{\pm m_{l}\atop \ \, n_{l}}$; and the rate constants of conformational changes between the open and closed structural states E and F by $k^{{\rm E}\to{\rm F}}_{m_{l}\atop n_{l}}$ and $k^{{\rm F}\to{\rm E}}_{m_{l}\atop n_{l}}$.  According to mass-action law, the rate of each reaction is given by the corresponding rate constant in the absence of incoming substrate, or by the rate constant multiplied by the concentration of the incoming substrate if any.  The kinetic equations of the reversible reaction network shown in Fig.~\ref{fig1} are given in Appendix~\ref{AppA}.  They constitute the master equations for the Markov jump stochastic process, ruling the time evolution of the probabilities of finding the DNA polymerase in any one of its structural states and the copy with a sequence $m_1m_2\cdots m_l$ of length~$l$.

\subsection{The kinetic equations}

\begin{figure}[h]
\includegraphics[width=15.5cm]{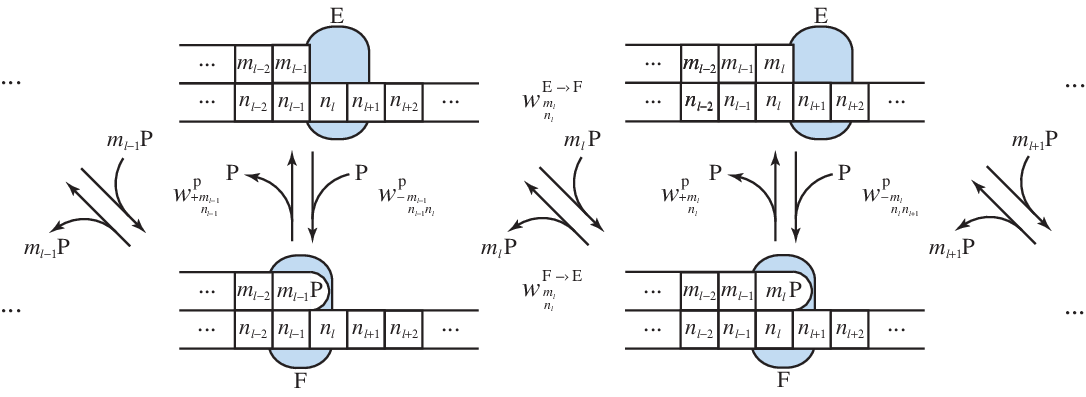}
\caption{Representation of the kinetic scheme of Fig.~\ref{fig1} after its reduction using the assumptions~(\ref{MM-conditions}) of a Michaelis-Menten enzyme kinetics for the DNA polymerase.  This reduced scheme corresponds to the kinetic equations~(\ref{kin_eq_E}) and~(\ref{kin_eq_F}), which are deduced in Appendix~\ref{AppA}. The notations are the same as in Fig.~\ref{fig1}. The transition rates are given by Eqs.~(\ref{w^E>F})-(\ref{w^p-}).}
\label{fig2}
\end{figure}

Now, according to Michaelis-Menten enzyme kinetics, we assume that the step 1 of substrate binding and unbinding is much faster than the other steps, i.e.,
\be
k_{+m\atop \ \, n} [m{\rm P}] \, , \ k_{-m\atop \ \, n} \  \gg \ k^{{\rm E}\to{\rm F}}_{m\atop n} \, , \ k^{{\rm F}\to{\rm E}}_{m\atop n} \, , \ k^{\rm p}_{+m\atop \ \, n}\, , \ k^{\rm p}_{-m\atop \ \, n} [{\rm P}] \, .
\label{MM-conditions}
\ee
Consequently, the reaction step 1 remains close to equilibrium during the process and, as shown in Appendix~\ref{AppA}, the kinetic equations simplify into
\bea
\frac{\dd}{\dd t}\, P_t(m_1 \cdots m_l ,l,{\rm E}) &=& w^{\rm p}_{+m_l\atop \ \, n_l} \, P_t(m_1 \cdots m_l ,l,{\rm F})
+\sum_{m_{l+1}} w^{{\rm F}\to{\rm E}}_{m_{l+1}\atop n_{l+1}}  \, P_t(m_1 \cdots m_l m_{l+1} ,l+1,{\rm F})
\nonumber\\
&&- \bigg( w^{\rm p}_{-m_l {\quad \ } \atop \ \ {n_l n_{l+1}}}  \!\! + \sum_{m_{l+1}} w^{{\rm E}\to{\rm F}}_{m_{l+1}\atop n_{l+1}} \bigg) P_t(m_1 \cdots m_l ,l,{\rm E}) \, ,
\label{kin_eq_E} \\
\frac{\dd}{\dd t}\, P_t(m_1 \cdots m_l ,l,{\rm F}) &=& w^{{\rm E}\to{\rm F}}_{m_l\atop n_l} \, P_t(m_1 \cdots m_{l-1} ,l-1,{\rm E})
+ w^{\rm p}_{-m_l {\quad \ }\atop \ \ n_l n_{l+1}} \! P_t(m_1 \cdots m_l ,l,{\rm E})
\nonumber\\
&&- \bigg( w^{\rm p}_{+m_l\atop \ \, n_l} + w^{{\rm F}\to{\rm E}}_{m_l\atop n_l} \bigg) P_t(m_1 \cdots m_l ,l,{\rm F}) \, ,
\label{kin_eq_F}
\eea
for the probabilities $P_t(m_1 \cdots m_l ,l,i)$ of finding the system at time $t$ in the states $(m_1 \cdots m_l ,l,i)$ with $m_j\in\{{\rm A},{\rm T}\}$ and $i\in\{{\rm E},{\rm F}\}$ at the location $l\in\{0,1,2,3,\dots\}$ along the template sequence $n_1n_2\cdots n_l n_{l+1}\cdots$.  The latter is supposed to remain invariant during the whole process.  The equations~(\ref{kin_eq_E}) and~(\ref{kin_eq_F}) satisfy the following condition of probability conservation
\be
\sum_{l=0}^{\infty} \sum_{i} \sum_{m_1\cdots m_l} P_t(m_1 \cdots m_l ,l,i) = 1
\label{tot-prob}
\ee
at every time $t$.  The calculations of Appendix~\ref{AppA} show that the transition rates are given by
\bea
w^{{\rm E}\to{\rm F}}_{m_l\atop n_l} &\equiv& k^{{\rm E}\to{\rm F}}_{m_l\atop n_l} \, \frac{ [ m_l{\rm P}]}{K_{m_l\atop n_l} Q_{n_l} } \, , \label{w^E>F}\\
w^{{\rm F}\to{\rm E}}_{m_l\atop n_l} &\equiv& k^{{\rm F}\to{\rm E}}_{m_l\atop n_l} \, , \label{w^F>E}\\
w^{\rm p}_{+m_l\atop \ \, n_l} &\equiv& k^{\rm p}_{+m_l\atop \ \, n_l} \, , \label{w^p+}\\
w^{\rm p}_{-m_l {\quad \ } \atop \ \ {n_l n_{l+1}}}  \!\!  &\equiv& k^{\rm p}_{-m_l \atop \ \, n_l} \, \frac{[{\rm P}]}{Q_{n_{l+1}}} \ , \label{w^p-}
\eea
with the Michaelis-Menten dissociation constants
\be
K_{m\atop n} \equiv \frac{k_{-m\atop \ \, n}}{k_{+m\atop \ \, n}}
\label{diss_csts}
\ee
and the denominators
\be
Q_n \equiv 1 + \sum_m \frac{[m{\rm P}]}{K_{m\atop n}} \, ,
\label{Q-dfn}
\ee
as expressed in terms of the rate constants introduced in Fig.~\ref{fig1}.  The reaction network corresponding to the reduced master equations~(\ref{kin_eq_E}) and~(\ref{kin_eq_F}) is shown in Fig.~\ref{fig2}.  We note that the depolymerization rates~(\ref{w^p-}) depend, not only on the pair $m_l\!:\!n_l$ that undergoes pyrophophorolysis, but also on the next template unit $n_{l+1}$, because the depolymerization steps are transitions from a state of quasi-equilibrium for the next pair $m_{l+1}\!:\!n_{l+1}$, as seen in Fig.~\ref{fig2}.

Moreover, we assume for simplicity that the constants of depolymerization are related to those of polymerization by
\be
k^{\rm p}_{-m\atop \ \, n}= \frac{1}{K_{\rm P}} \, k^{\rm p}_{+m\atop \ \, n}
\label{K_P-dfn}
\ee
with a unique constant of pyrophosphorolysis $K_{\rm P}$.

The process of Fig.~\ref{fig2} starts from an initial state, which may be taken as $(\emptyset,0,{\rm E})$ with the polymerase in its open structural state E and a copy of length $l=0$.  Initially, there is thus no possibility of depolymerization, so that we should have $w^{\rm p}_{-\emptyset {\quad \ \,} \atop \ \ {\emptyset n_{l+1}}}=0$.  Accordingly, there are only two possible initiation steps: the attachment of either the unit $m_1={\rm A}$ or $m_1={\rm T}$ with the transition ${\rm E}\to{\rm F}$ at the rate $w^{{\rm E}\to{\rm F}}_{m_1\atop n_1}$.

\subsection{The case of the T7 DNA polymerase}
\label{subsec:T7_DNA_pol}

We apply the model to the DNA polymerase of the bacteriophage T7, for which experimental measurements have given ranges of values for the rate constants \cite{DJ20,DJ21,DKJ22}.  In order to investigate numerically the process, we need to give specific values to the rate constants.  As discussed in Appendix~\ref{AppB}, we here adopt the values of Table~\ref{tab:pol-param}, which are consistent with the experimental data of Refs.~\cite{DJ20,DJ21,DKJ22}.

In consistency with these data, the pyrophosphorolysis constant should take the value
\be
K_{\rm P} = 5.6 \times 10^{-3} \, {\rm M} \, ,
\label{K_P-value}
\ee
as explained in Appendix~\ref{AppB}.  Furthermore, the pyrophosphate concentration is assumed to be equal to $[{\rm P}]=10^{-4}\, {\rm M}$ \cite{H01}.

\begin{table}[h]
\caption{Kinetic parameters for the two-state model of T7 DNA polymerase: The values of the constants used in computations are adopted following the discussion of Appendix~\ref{AppB} based on the experimental data of Refs.~\cite{DJ20,DJ21,DKJ22}.  $K_{m\atop n}$ denote the dissociation constants~(\ref{diss_csts}).  The rate constants $k^{{\rm E}\to{\rm F}}_{m\atop n}$, $k^{{\rm F}\to{\rm E}}_{m\atop n}$, and $k^{\rm p}_{+m\atop \ \, n}$ are those introduced in Fig.~\ref{fig1}.}
\label{tab:pol-param}
\vspace{5mm}
\begin{center}
\begin{tabular}{|cc|cccc|}
\hline
$m{\rm P}$ & $n$ & $K_{m\atop n}$ (M) & \ $k^{{\rm E}\to{\rm F}}_{m\atop n}$ (s$^{-1}$) & \ $k^{{\rm F}\to{\rm E}}_{m\atop n}$ (s$^{-1}$) & \ $k^{\rm p}_{+m\atop \ \, n}$ (s$^{-1}$)\\
\hline
dATP & A &\ $2.07\times 10^{-2}$ & $170$ & $340$ & $6.2$ \\
dATP & T &\ $4.15\times 10^{-4}$ & $6500$ & $1.7$ & $293$ \\
dTTP & A &\ $3.92\times 10^{-4}$ & $6500$ & $1.7$ & $311$ \\
dTTP & T &\ $9.3\times 10^{-3}$ & $170$ & $340$ & $2.4$ \\
\hline
\end{tabular}
\end{center}
\end{table} 

\subsection{Simulating the process and comparing with theory}

 On the one hand, the process in Fig.~\ref{fig2} can be numerically simulated with Gillespie's numerical algorithm \cite{G76,G77}.  This algorithm generate random trajectories, which are sampled at the discrete times of the jumps between successive states for the system.  If the copy has the length $l$ and the polymerase is in the state E, there are three possible jumps: 1) the attachment of the copy unit $m_{l+1}={\rm A}$ with the transition ${\rm E}\to{\rm F}$ at the rate $w^{{\rm E}\to{\rm F}}_{m_{l+1}\atop n_{l+1}}$ depending on the template unit $n_{l+1}$; 2) the attachment of the copy unit $m_{l+1}={\rm T}$ with the transition ${\rm E}\to{\rm F}$ at the rate $w^{{\rm E}\to{\rm F}}_{m_{l+1}\atop n_{l+1}}$ also depending on the template unit $n_{l+1}$; and 3) the transition ${\rm E}\to{\rm F}$ at the rate $w^{\rm p}_{-m_l {\quad \ } \atop \ \ {n_l n_{l+1}}}$ without change of the length~$l$ and depending on the pair $m_l\!:\! n_l$ and the template unit $n_{l+1}$.  If the copy has the length $l$ and the polymerase is in the state F, there are two possible jumps: 1) the transition ${\rm F}\to{\rm E}$ at the rate $w^{\rm p}_{+m_l\atop \ \, n_l}$ without change of the length~$l$ and depending on the pair $m_l\!:\! n_l$; and 2) the detachment of the previously incorporated unit, which is either $m_l={\rm A}$ or $m_l={\rm T}$, at the rate $w^{{\rm F}\to{\rm E}}_{m_l\atop n_l}$ depending on the pair $m_l\!:\!n_l$.  The random waiting times $\Delta t$ before each jump are exponentially distributed with the probability density $p(\Delta t)={\mathcal T}^{-1} \exp(-\Delta t/{\mathcal T})$, where the parameter ${\mathcal T}$ is given by the inverse of the sum of the rates for the possible jumps.

On the other hand, as shown in the following Section~\ref{sec:theory}, exact formulas are known for the mean grown velocity, the mean replication error, the local probabilities of finding correct or incorrect base pairs, and the mean dwell times of the polymerase at the successive locations along the template, according to the theory developed in the companion paper \cite{paperI}.  Our goal is to compare the two methods.

\section{Theory}
\label{sec:theory}

\subsection{Matrix form for the kinetic equations}

As explained in the companion paper~\cite{paperI}, since the polymerase has $I=2$ structural states, we here introduce the $2\times 2$ probability matrices,
\be
\boldsymbol{\mathsf P}_t(m_1\cdots m_l,l) \equiv
\left[
\begin{array}{cc}
P_t(m_1\cdots m_l,l,{\rm E}) & P_t(m_1\cdots m_l,l,{\rm E}) \\
P_t(m_1\cdots m_l,l,{\rm F}) & P_t(m_1\cdots m_l,l,{\rm F}) \\
\end{array}
\right] ,
\label{matrix-P}
\ee
which are formed with two identical columns containing the probabilities $P_t(m_1 \cdots m_l ,l,i)$ of finding the polymerase in its states $i\in\{{\rm E},{\rm F}\}$ and the copy with the sequence $m_1 \cdots m_l$ at time $t$.  Hence, the master equations~(\ref{kin_eq_E}) and~(\ref{kin_eq_F}) can be expressed in matrix form as
\bea
\frac{\dd}{\dd t}\, \boldsymbol{\mathsf P}_t(m_1\cdots m_l,l) &=& \boldsymbol{\mathsf W}_{+m_l,l}^{\rm c} \cdot \boldsymbol{\mathsf P}_t(m_1 \cdots m_{l-1} ,l-1)
 + \sum_{m_{l+1}} \boldsymbol{\mathsf W}_{-m_{l+1},l+1}^{\rm c} \cdot  \boldsymbol{\mathsf P}_t(m_1 \cdots m_l m_{l+1} ,l+1)
\nonumber\\
&&+ \, \bigg(\boldsymbol{\mathsf W}_{m_l,l}^{0} - \boldsymbol{\mathsf W}_{-m_l,l}^{\rm d} - \sum_{m_{l+1}} \boldsymbol{\mathsf W}_{+m_{l+1},l+1}^{\rm d} \bigg) \cdot\boldsymbol{\mathsf P}_t(m_1 \cdots m_l ,l)
\label{matrix_kin_eq}
\eea
in terms of the following $2\times 2$ matrices containing the transition rates~(\ref{w^E>F})-(\ref{w^p-}):
\be
\boldsymbol{\mathsf W}_{+m_l,l}^{\rm c} \equiv
\left[
\begin{array}{cc}
0 & 0  \\
w^{{\rm E}\to{\rm F}}_{m_l\atop n_l} & 0  \\
\end{array}
\right] ,
\qquad
\boldsymbol{\mathsf W}_{-m_{l+1},l+1}^{\rm c} \equiv
\left[
\begin{array}{cc}
0 & w^{{\rm F}\to{\rm E}}_{m_{l+1}\atop n_{l+1}} \\
0 & 0  \\
\end{array}
\right] ,
\label{matrix-Wc}
\ee
\be
\boldsymbol{\mathsf W}_{+m_{l+1},l+1}^{\rm d} \equiv
\left[
\begin{array}{cc}
w^{{\rm E}\to{\rm F}}_{m_{l+1}\atop n_{l+1}} & 0 \\
0 & 0\\
\end{array}
\right] ,
\qquad
\boldsymbol{\mathsf W}_{-m_l,l}^{\rm d} \equiv
\left[
\begin{array}{cc}
0 & 0 \\
0 & w^{{\rm F}\to{\rm E}}_{m_l\atop n_l} \\
\end{array}
\right] ,
\label{matrix-Wd}
\ee
and
\be
\boldsymbol{\mathsf W}_{m_l,l}^{0} \equiv
\left[
\begin{array}{cc}
- w^{\rm p}_{-m_l {\quad \ } \atop \ \ {n_l n_{l+1}}} & w^{\rm p}_{+m_l\atop \ \, n_l} \\
w^{\rm p}_{-m_l {\quad \ } \atop \ \ {n_l n_{l+1}}}   & - w^{\rm p}_{+m_l\atop \ \, n_l} \\
\end{array}
\right] .
\label{W0-matrix}
\ee

\subsection{Backward and forward iterations}

According to the theory of the companion paper~\cite{paperI}, the exact asymptotic solution of the master equations will be given by the {\it backward iteration}:
\be
\boldsymbol{\mathsf V}_{l-1}= \sum_{m_l} \boldsymbol{\mathsf W}^{\rm d}_{+m_l,l} - \sum_{m_l} \boldsymbol{\mathsf W}^{\rm c}_{-m_l,l} \cdot \left( \boldsymbol{\mathsf V}_l - \boldsymbol{\mathsf W}^{0}_{m_l,l} + \boldsymbol{\mathsf W}^{\rm d}_{-m_l,l}\right)^{-1} \cdot \boldsymbol{\mathsf W}^{\rm c}_{+m_l,l} \, ,
\label{backward_iter}
\ee
running along the template sequence $n_1n_2\cdots n_l\cdots n_L$ to obtain the $2\times 2$ matrices $\boldsymbol{\mathsf V}_l$ and, thus, also the following $2\times 2$ matrices,
\be
\boldsymbol{\mathsf Y}_{m_l,l} = \left( \boldsymbol{\mathsf V}_l - \boldsymbol{\mathsf W}^{0}_{m_l,l} + \boldsymbol{\mathsf W}^{\rm d}_{-m_l,l}\right)^{-1} \cdot \boldsymbol{\mathsf W}^{\rm c}_{+m_l,l} \, ,
\label{Y_ml}
\ee
which determine the local composition of the copy.  Furthermore, there is the {\it forward iteration}:
\be
\boldsymbol{\Psi}_l = \boldsymbol{\mathsf R}_l \cdot \boldsymbol{\Psi}_{l-1}
\qquad\mbox{with}\qquad
\boldsymbol{\mathsf R}_l \equiv \sum_{m_l} \boldsymbol{\mathsf Y}_{m_l,l} \, ,
\label{forward_iter}
\ee
in order to calculate the $2\times 2$ matrices $\boldsymbol{\Psi}_l$. 

\subsection{The observable quantities}
\label{subsec:observables}

Once, these sets of matrices are determined, the {\it mean local dwell time} of the polymerase at the location~$l$ of the template is given by
\be
\tau_l = \frac{1}{C} \, {\rm tr} \, \boldsymbol{\Psi}_l
\qquad\mbox{with}\qquad
C= {\rm tr}(\boldsymbol{\mathsf V}_l \cdot \boldsymbol{\Psi}_l) ={\rm tr}(\boldsymbol{\mathsf V}_{l-1} \cdot \boldsymbol{\Psi}_{l-1})
\label{tau_l}
\ee
and the {\it mean growth velocity} by
\be
v = \frac{1}{\langle \tau_l \rangle}
\qquad\mbox{with}\qquad
\langle\tau_l \rangle \equiv \lim_{L\to\infty} \frac{1}{L} \sum_{l=1}^L \tau_l \, ,
\label{v-formula}
\ee
after averaging the mean local dwell times over the template sequence.

The {\it sequence probabilities}, i.e., the probabilities of finding a given sequence for the copy, can thus be expressed as
\be
\mu(m_1\cdots m_L;L) = \frac{1}{{\rm tr}\, \boldsymbol{\Psi}_L} \, {\rm tr}(\boldsymbol{\mathsf Y}_{m_L,L} \cdots \boldsymbol{\mathsf Y}_{m_1,1} \cdot \boldsymbol{\Psi}_0)
\label{mu(seq)-formula}
\ee
in terms of the matrices~(\ref{Y_ml}) and the initial condition ${\Psi}_0$ of the forward iteration~(\ref{forward_iter}).  Consequently, the {\it local probabilities} of finding the units $m=m_l\in\{{\rm A},{\rm T}\}$ at the location~$l$ of the copy can be calculated with
\be
\mu(m;l) = \lim_{L\to\infty} \frac{1}{{\rm tr}\, \boldsymbol{\Psi}_L} \, {\rm tr}(\boldsymbol{\mathsf S}_l \cdot \boldsymbol{\mathsf Y}_{m,l}  \cdot \boldsymbol{\Psi}_{l-1})
\qquad\mbox{with}\qquad
\boldsymbol{\mathsf S}_l \equiv \boldsymbol{\mathsf R}_L \cdots \boldsymbol{\mathsf R}_{l+1} \, .
\label{mu(m;l)-formula}
\ee
Accordingly, the {\it error probability} of the replication process by the DNA polymerase is found to be equal to
\be
\eta \equiv \lim_{L\to\infty} \frac{1}{L} \sum_{l=1}^L \left[ 1 - \mu(\tilde n_l;l)\right] ,
\label{eta-dfn}
\ee
where $\tilde n_l$ denotes the copy unit that is complementary to the template unit $n_l$ at the location $l$ of the sequence.  Since we have assumed that the template and the copy have only two types of nucleotides (A and T), the correct pairs are A:T and T:A and the incorrect pairs A:A and T:T, whereupon the incorrect pairs causing replication errors are the pairs $n_l\!:\!n_l$ with the same unit at the location~$l$ in the copy and the template.  Under such circumstances, the replication error~(\ref{eta-dfn}) can be written as
\be
\eta \equiv \lim_{L\to\infty} \frac{1}{L} \sum_{l=1}^L \mu(n_l;l)
\label{eta-formula}
\ee
in terms of the local probability~(\ref{mu(m;l)-formula}) for $m=n_l$.

\subsection{Reduction from matrices to scalars}

We note that the $2\times 2$ matrices~(\ref{matrix-Wc}) and~(\ref{matrix-Wd}) all have a column with zeros.  As a consequence, there is a reduction from of the matrix iterations~(\ref{backward_iter}) and~(\ref{forward_iter}) into scalar iterations.  Indeed, if we suppose that the matrix $\boldsymbol{\mathsf V}_l$ has the following form,
\be
\boldsymbol{\mathsf V}_l = 
\left[
\begin{array}{cc}
x_l & 0 \\
0 & 0 \\
\end{array}
\right] ,
\label{V_l-x_l}
\ee
then the matrix $\boldsymbol{\mathsf V}_{l-1}$ given by the backward iteration~(\ref{backward_iter}) has again the same form with $x_{l-1}$ replacing $x_l$.  Therefore, the backward iteration~(\ref{backward_iter}) reduces to the following scalar iteration:
\be
x_{l-1} = f_{n_l n_{l+1}}(x_l) 
\qquad\mbox{with}\qquad
f_{n_l n_{l+1}}(x) \equiv x \sum_m \frac{\alpha_{m\atop n_l}}{x+\beta_{m {\quad \ \ } \atop {n_l n_{l+1}}}} \, ,
\label{IFS}
\ee
with
\bea
\alpha_{m\atop \, n_l} &\equiv& \frac{w^{\rm p}_{+m\atop \ \ n_l}  \, w^{{\rm E}\to{\rm F}}_{m\atop \, n_l}}{w^{\rm p}_{+m\atop \ \ n_l} + w^{{\rm F}\to{\rm E}}_{m\atop \, n_l}} \, , \label{alpha}\\
\beta_{m {\quad \ \ } \atop {n_l n_{l+1}}} &\equiv& \frac{w^{\rm p}_{-m {\quad \ \ } \atop \ \ {n_l n_{l+1}}} w^{{\rm F}\to{\rm E}}_{m\atop \, n_l}}{w^{\rm p}_{+m\atop \ \ n_l} + w^{{\rm F}\to{\rm E}}_{m\atop \, n_l}} \, , \label{beta}
\eea
which runs backward along the template sequence and is determined by the successive doublets $n_l n_{l+1}$ of template units.  The backward iteration~(\ref{IFS}) forms an {\it iterated function system} \cite{BD85}, as considered in Refs.~\cite{G16PRL,G17PRE} for the kinetics of single-state DNA polymerases.

Similarly, the $2\times 2$ matrices~(\ref{Y_ml}) have the form,
\be
\boldsymbol{\mathsf Y}_{m_l,l} = 
\left[
\begin{array}{cc}
Y^{\rm E}_{m_l,l} & 0 \\
Y^{\rm F}_{m_l,l} & 0 \\
\end{array}
\right]
\label{Y_ml-2}
\ee
with
\be
Y^{\rm E}_{m_l,l} \equiv \frac{\alpha_{m_l\atop \, n_l}}{x_l+\beta_{m_l {\quad \ \, } \atop {n_l n_{l+1}}}}
\qquad\mbox{and}\qquad
Y^{\rm F}_{m_l,l} \equiv
\frac{\alpha_{m_l\atop \, n_l}}{w^{\rm p}_{+m_l\atop \ \, n_l}} \, \frac{x_l+ w^{\rm p}_{-m_l {\quad \ \, } \atop \ \ {n_l n_{l+1}}} }{x_l+\beta_{m_l {\quad \ \, } \atop {n_l n_{l+1}}}} \, ,
\label{YE-YF}
\ee
and their second column filled with zeros.  Therefore, the forward iteration~(\ref{forward_iter}) is ruled by $2\times 2$ matrices having the same form:
\be
\boldsymbol{\mathsf R}_l = 
\left[
\begin{array}{cc}
R^{\rm E}_l & 0 \\
R^{\rm F}_l & 0 \\
\end{array}
\right] ,
\label{R_l-2}
\ee
where
\be
R^{\rm E}_l = \sum_{m_l} Y^{\rm E}_{m_l,l} = \frac{x_{l-1}}{x_l} \, ,
\label{RE_l}
\ee
as a consequence of Eqs.~(\ref{IFS}) and~(\ref{YE-YF}), and
\be
R^{\rm F}_l = \sum_{m_l} Y^{\rm F}_{m_l,l} =   \sum_{m_l} \frac{\alpha_{m_l\atop \, n_l}}{w^{\rm p}_{+m_l\atop \ \, n_l}} \, \frac{x_l+ w^{\rm p}_{-m_l {\quad \ \, } \atop \ \ {n_l n_{l+1}}} }{x_l+\beta_{m_l {\quad \ \, } \atop {n_l n_{l+1}}}} \, .
\label{RF_l}
\ee
Because of the form~(\ref{R_l-2}), the forward iteration~(\ref{forward_iter}) implies that
\be
\boldsymbol{\Psi}_l = 
\left[
\begin{array}{cc}
\psi^{\rm E}_l & \psi^{\rm E}_l \\
\psi^{\rm F}_l & \psi^{\rm F}_l \\
\end{array}
\right]
\qquad\mbox{with}\qquad
\psi^{\rm E}_l  = R^{\rm E}_l  \, \psi^{\rm E}_{l-1}
\qquad\mbox{and}\qquad
\psi^{\rm F}_l  = R^{\rm F}_l  \, \psi^{\rm E}_{l-1} \, .
\label{Psi_l-2}
\ee
Furthermore, the matrix defined in Eq.~(\ref{mu(m;l)-formula}) is now given by
\be
\boldsymbol{\mathsf S}_l = R^{\rm E}_L R^{\rm E}_{L-1} \cdots R^{\rm E}_{l+1} \left[
\begin{array}{cc}
1 & 0 \\
\frac{R^{\rm F}_L}{R^{\rm E}_L} & 0 \\
\end{array}
\right] .
\label{S_l-2}
\ee

For $l=0$, we obtain the $2\times 2$ matrix $\boldsymbol{\mathsf S}_0$.  If we assume that the template sequence is periodic of period $L$, the replication process can be envisaged as running on a loop with periodic boundary conditions in the growth regime.  In this case, we should have that $\boldsymbol{\Psi}_{l+L}= \boldsymbol{\Psi}_l$ for $l=1,2,\dots,L$ and $\boldsymbol{\mathsf S}_0\cdot\boldsymbol{\Psi}_L=\boldsymbol{\Psi}_L$, so that $\prod_{l=1}^L R^{\rm E}_l=1$.  Thus, we find that
\be
\psi^{\rm E}_L = \frac{R^{\rm E}_L}{R^{\rm E}_L + R^{\rm F}_L}
\qquad\mbox{and}\qquad
\psi^{\rm F}_L = \frac{R^{\rm F}_L}{R^{\rm E}_L + R^{\rm F}_L} \, .
\label{PsiEF_L}
\ee

The constant $C$ in Eq.~(\ref{tau_l}) is thus equal to $C=x_l\psi^{\rm E}_l =x_{l-1}\psi^{\rm E}_{l-1}$, so that $\psi^{\rm E}_l =C/x_l$.  Therefore, the mean local dwell time of the polymerase at the location~$l$ of the template is given by
\be
\tau_l= \frac{1}{C} \, \left(\psi^{\rm E}_l + \psi^{\rm F}_l \right) = \frac{1}{x_l} \left( 1 + \frac{R^{\rm F}_l}{R^{\rm E}_l} \right) ,
\label{tau_l-scalar}
\ee
since we have that $\psi^{\rm F}_l/\psi^{\rm E}_l=R^{\rm F}_l/R^{\rm E}_l$ because of Eq.~(\ref{Psi_l-2}).
Therefore, the mean growth velocity is given by
\be
\frac{1}{v} = \left\langle \frac{1}{x_l} \left( 1 + \frac{R^{\rm F}_l}{R^{\rm E}_l} \right)\right\rangle ,
\label{v-scalar}
\ee
where $\langle\cdot\rangle$ denotes the average over the template sequence.

The sequence probabilities~(\ref{mu(seq)-formula}) thus become
 \be
\mu(m_1\cdots m_L;L) = \frac{Y^{\rm E}_{m_L,L}+Y^{\rm F}_{m_L,L}}{R^{\rm E}_L+R^{\rm F}_L}  \prod_{l=1}^{L-1} \frac{Y^{\rm E}_{m_l,l}}{R^{\rm E}_l}\, , 
\label{mu(seq)-scalar}
\ee
satisfying the normalization condition $\sum_{m_1\cdots m_L} \mu(m_1\cdots m_L;L) =1$.
Accordingly, the local probabilities~(\ref{mu(m;l)-formula}) are given by
\be
\mu(m;l) = \frac{Y^{\rm E}_{m,l}}{R^{\rm E}_l} = \frac{x_l}{x_{l-1}} \, \frac{\alpha_{m\atop n_l}}{x+\beta_{m {\quad \ \ } \atop {n_l n_{l+1}}}}
\label{mu(m;l)-scalar}
\ee
and the error probability~(\ref{eta-formula}) by
\be
\eta = \left\langle \frac{Y^{\rm E}_{n_l,l}}{R^{\rm E}_l} \right\rangle ,
\label{eta-scalar}
\ee
after averaging over the template sequence.  

Therefore, theory provides exact formulas for the quantities of interest.  We note that the reduction from matrix to scalar form is possible because the reaction network~(\ref{pol-react}) is strictly sequential, as seen in Figs.~\ref{fig1} and~\ref{fig2}.

\subsection{The different regimes of the process}

As discussed in the companion paper~\cite{paperI}, the process of template-directed copolymerization catalyzed by the DNA polymerase may run in different regimes, depending on the nucleotide concentrations with respect to the pyrophosphate concentration and on the template sequence.

\subsubsection{Stationary regime}

If the nucleotide concentrations are too small, the growth of the copy is stalled.  In this stationary regime, the mean length of the copy remains finite up to nucleotide concentration values corresponding to the onset of growth, as determined in Appendix~\ref{App:Sublin}.
For larger values of the nucleotide concentrations, the growth of the copy becomes possible.

\subsubsection{Regime of sublinear growth in time}

As explained in Appendix~\ref{App:Sublin}, if the template sequence is disordered, there is a regime where the mean length has a sublinear growth in time as $\langle l\rangle_t \sim t^\gamma$ with an exponent $0 < \gamma < 1$, where $\gamma=0$ at the onset of growth and $\gamma=1$ at the threshold of the steady-growth regime.

\subsubsection{Steady-growth regime}

In this regime, the mean length of the copy grows linearly in time like $\langle l\rangle_t \simeq vt$ with a positive mean growth velocity $v>0$ given by Eq.~(\ref{v-scalar}).  Close to the threshold $\gamma=1$, the depolymerization rates $w^{\rm p}_{-m_l {\quad \ } \atop \ \ {n_l n_{l+1}}}$ continue to play a significant role in determining the velocity $v$.  However, their role becomes negligible away beyond the threshold of steady growth, when the process enters into another regime, where the step~3 of the reactions~(\ref{pol-react}), i.e., the release of pyrophosphate, can be assumed to be quasi irreversible.

Next, because of the Michaelis-Menten kinetics of the DNA polymerase, there is a crossover between two different regimes of elongation, depending on the values of the nucleotide concentrations with respect to the dissociation constants~(\ref{diss_csts}).

If the nucleotide concentrations are lower than the dissociation constants, the elongation of the copy is controlled by the arrival of substrates (i.e., nucleotides) in the polymerase, which corresponds to the step 1 of reactions~(\ref{pol-react}). Otherwise, the elongation is controlled by the polymerization reaction and the release of pyrophosphate in the last two steps 2 and 3 of reactions~(\ref{pol-react}) and the polymerase reaches its full speed along the template.

\subsection{The irreversible approximation}

If the nucleotide concentrations are sufficiently larger than their values at the threshold $\gamma=1$ of steady growth, the detachment rates may be assumed to be negligible with respect to the other rates, i.e., $w^{\rm p}_{-m_l {\quad \ } \atop \ \ {n_l n_{l+1}}} \!\!\! =0$. Therefore, the parameters~(\ref{beta}) can also be neglected, i.e., $\beta_{m {\quad \ \ } \atop {n_l n_{l+1}}} \!\!\! =0$.  In this case, the iterated function system~(\ref{IFS}) can be expressed as
\be
x_{l-1} = \sum_m \alpha_{m\atop n_l}
\label{IFS-0}
\ee
in terms of the parameters~(\ref{alpha}).  Accordingly, in the irreversible approximation, the process becomes independent of the next template unit $n_{l+1}$.  We thus have
\be
Y^{\rm E}_{m_l,l} \equiv \frac{\alpha_{m_l\atop \, n_l}}{x_l}
\qquad\mbox{and}\qquad
Y^{\rm F}_{m_l,l} \equiv
\frac{\alpha_{m_l\atop \, n_l}}{w^{\rm p}_{+m_l\atop \ \, n_l}} \, ,
\label{YE-YF-0}
\ee
so that the local probabilities are given by
\be
\mu(m;l) = \frac{\alpha_{m\atop \, n_l}}{\sum_{m^\prime} \alpha_{m^\prime\atop n_l}}
\label{mu(m;l)-scalar-0}
\ee
and the error probability by
\be
\eta = \left\langle \frac{\alpha_{n_l\atop \, n_l}}{\sum_{m} \alpha_{m\atop \, n_l}} \right\rangle .
\label{eta-scalar-0}
\ee
In the regime where this irreversible approximation holds, the mean growth velocity increases with the nucleotide concentrations if they are smaller than the dissociation constants~(\ref{diss_csts}), but the velocity reaches a plateau where the polymerase runs at its maximum velocity if the nucleotide concentrations are larger than the dissociation constants~(\ref{diss_csts}), which is the behavior expected for a Michaelis-Menten enzyme kinetics.

\section{Numerical results}
\label{sec:results}

We consider the process of multistate template-directed copolymerization by the T7 DNA polymerase with the kinetic parameters of Subsection~\ref{subsec:T7_DNA_pol}.  The template sequence is taken as a Bernoulli chain with probabilities $\nu({\rm A})$ and $\nu({\rm T})=1-\nu({\rm A})$.  We study the observable quantities of Subsection~\ref{subsec:observables} for the process of interest.  On the one hand, they are computed using numerical simulation with Gillespie's algorithm~\cite{G76,G77}, which requires performing a statistics over a large enough number $N_{\rm stat}$ of random copies.  On the other hand, these quantities are also computed using the theoretical formulas of Section~\ref{sec:theory}, which requires convergence over some number $N_{\rm loop}$ of loops formed by a template sequence of length $L$.  Since both methods are exact, the results should coincide, which we check.  Moreover, we compare the amounts of computer time it takes to run computations with the two methods.

\subsection{Mean growth velocity and error probability}

According to theory, the mean growth velocity is given by Eq.~(\ref{v-scalar}) and the error probability by Eq.~(\ref{eta-scalar}).

\begin{figure}[h]
\includegraphics[width=9cm]{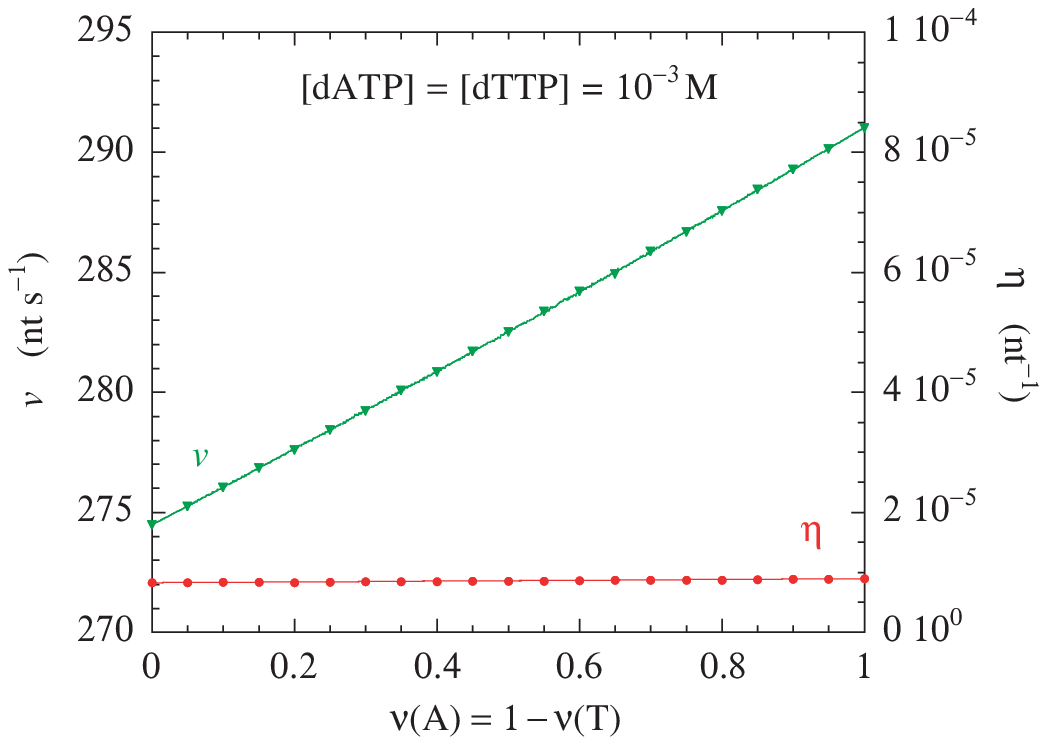}
\caption{The mean growth velocity $v$ and the error probability $\eta$ versus the fraction $\nu({\rm A})$ of units A in the Bernoulli chain forming the template sequence of length $L=10^5$ for the nucleotide concentrations $[{\rm dATP}]=[{\rm dTTP}]=10^{-3}$~M. The dots are the results of Gillespie's numerical algorithm using a statistics over $N_{\rm stat}=10^5$ copies.  The lines are those of theory using the formulas~(\ref{v-scalar}) for the velocity and~(\ref{eta-scalar}) for the error probability with $N_{\rm loop}=10$.}
\label{fig3}
\end{figure}

Figure~\ref{fig3} shows the mean growth velocity and the error probability of the replication process as a function of the fraction $\nu({\rm A})=1-\nu({\rm T})$ of bases A in the random template sequence for fixed and equal nucleotide concentrations.  In this case, the dependence of both quantities can be fitted to the following linear functions:
\bea
[{\rm dATP}]=[{\rm dTTP}]=10^{-3}\, {\rm M}: \qquad && v = 291.02 \ {\rm nt} \, {\rm s}^{-1} \times \nu({\rm A}) + 274.47 \ {\rm nt} \, {\rm s}^{-1} \times \nu({\rm T}) \, , \\
&& \eta = 8.9182 \times 10^{-6} \ {\rm nt}^{-1} \times \nu({\rm A}) + 8.2279 \times 10^{-6} \ {\rm nt}^{-1} \times \nu({\rm T}) \, ,
\eea
where $\nu({\rm T})=1-\nu({\rm A})$.  These results show that the T7 DNA polymerase runs fast and has a high fidelity along random templates.  The polymerase is slightly faster if the template contains more A than T.

\begin{figure}[h]
\includegraphics[width=9cm]{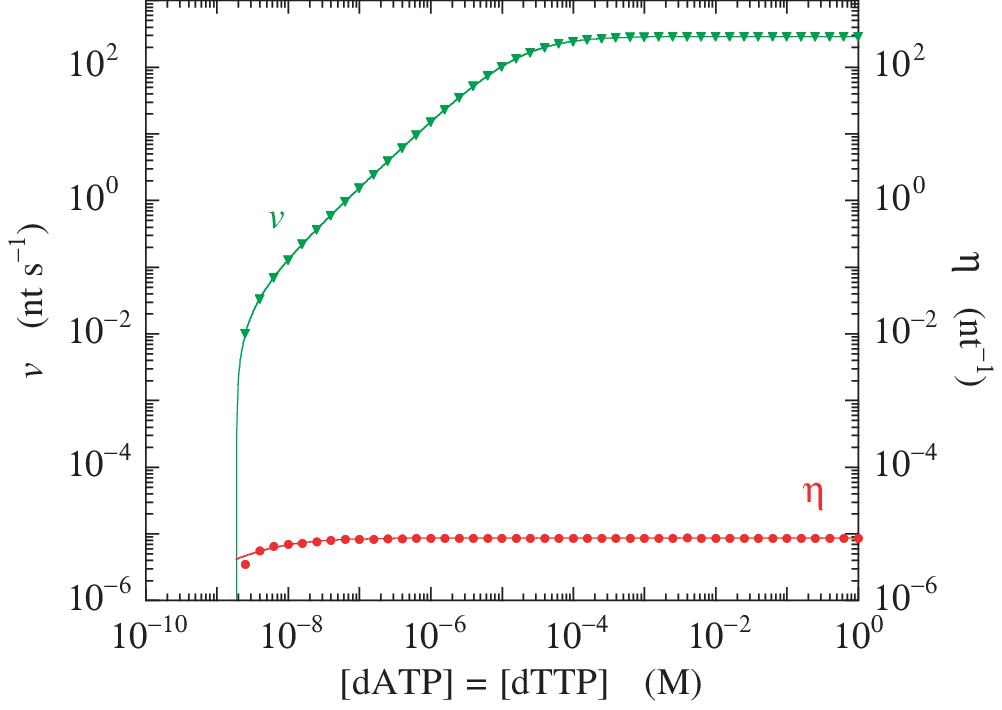}
\caption{The mean growth velocity $v$ and the error probability $\eta$ versus equal values for the nucleotide concentrations $[{\rm dATP}]=[{\rm dTTP}]$.  Here, the template sequence is a Bernoulli chain with probabilities $\nu({\rm A})=\nu({\rm T})=0.5$ and length $L=10^5$. The dots are the results of Gillespie's numerical algorithm using a statistics over $N_{\rm stat}=10^5$ copies.  The lines are those of theory using the formulas~(\ref{v-scalar}) for the velocity and~(\ref{eta-scalar}) for the error probability with $N_{\rm loop}=10$.}
\label{fig4}
\end{figure}

In the following, we consider random template sequences with equal fractions of both bases: $\nu({\rm A})=\nu({\rm T})=0.5$.

Figure~\ref{fig4} depicts the mean growth velocity and the error probability now as a function of equal nucleotide concentrations, $[{\rm dATP}]=[{\rm dTTP}]$.  We observe therein that the mean growth velocity vanishes at the concentration value:
\be
v=0: \qquad [{\rm dATP}]=[{\rm dTTP}] = 1.8846 \times 10^{-9} \, {\rm M} \, ,
\label{dNTP-v=0}
\ee
which agrees with the nucleotide concentration value~(\ref{A=T,g=1}) for the threshold of linear growth in time, as expected.
Next, up to concentrations of about $10^{-5}$-$10^{-4}$~M, the mean growth velocity increases linearly with the nucleotide concentrations, because the overall rate of the polymerase is controlled under these conditions by the arrival of nucleotides.
Beyond, the polymerase runs at full speed and the mean growth velocity reaches a plateau, as well as the error probability:
\bea
\mbox{full-speed regime}: \qquad && v = 287.652 \ {\rm nt} \, {\rm s}^{-1} \, , \\
&& \eta = 8.5726 \times 10^{-6} \ {\rm nt}^{-1} \, ,
\eea
for $[{\rm dATP}]=[{\rm dTTP}]=1~{\rm M}$.  The behavior observed in Fig.~\ref{fig4} is characteristic of a Michaelis-Menten enzyme kinetics.  Indeed, for $[{\rm dNTP}]\equiv[{\rm dATP}]=[{\rm dTTP}] \gtrsim 10^{-6}~{\rm M}$, the velocity can be fitted to
\be
v = \frac{k_{\rm cat}\, [{\rm dNTP}]}{[{\rm dNTP}] + K_{\rm m}}
\qquad\mbox{with}\qquad
k_{\rm cat} = 287.9(3) \ {\rm nt} \, {\rm s}^{-1}
\qquad\mbox{and}\qquad
K_{\rm m} = 1.832(4) \times 10^{-5} \, {\rm M} \, ,
\label{v-MM}
\ee
where the number in parentheses is the standard deviation uncertainty in the last digit of the quoted value for both parameters.

\begin{figure}[h]
\includegraphics[width=9cm]{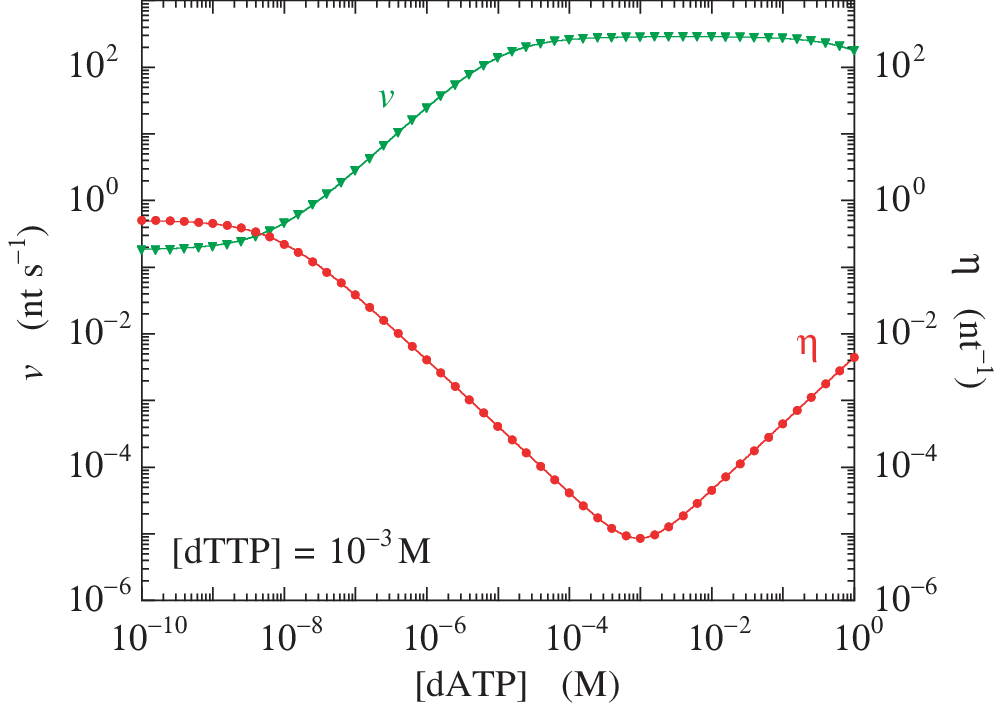}
\caption{The mean growth velocity $v$ and the error probability $\eta$ versus the nucleotide concentration $[{\rm dATP}]$ for the fixed value $[{\rm dTTP}]=10^{-3}$~M of the other nucleotide.  The template sequence is a Bernoulli chain with probabilities $\nu({\rm A})=\nu({\rm T})=0.5$ and length $L=10^5$. The dots are the results of Gillespie's numerical algorithm using a statistics over $N_{\rm stat}=10^5$ copies.  The lines are those of theory using the formulas~(\ref{v-scalar}) for the velocity and~(\ref{eta-scalar}) for the error probability with $N_{\rm loop}=10$.}
\label{fig5}
\end{figure}

Figure~\ref{fig5} shows the same quantities as in Fig.~\ref{fig4}, but as a function of the nucleotide concentration $[{\rm dATP}]$ for a fixed value of the other nucleotide concentration $[{\rm dTTP}]$. In Fig.~\ref{fig5}, we observe that the mean growth velocity reaches a maximum value and the error probability a minimum value for
\bea
[{\rm dTTP}]=10^{-3}~{\rm M}: \qquad && v_{\rm max} = 284.255 \ {\rm nt} \, {\rm s}^{-1}\qquad \qquad\mbox{at}\qquad  [{\rm dATP}]=4.17 \times 10^{-3}~{\rm M}\, , \\
&& \eta_{\rm min} = 8.5663 \times 10^{-6} \ {\rm nt}^{-1} \qquad\mbox{at}\qquad  [{\rm dATP}]=0.95 \times 10^{-3}~{\rm M}\, ,
\eea
i.e., for the two nucleotide concentrations having the same order of magnitude.  This means that the polymerase is most efficient to achieve a high fidelity, if the pool of nucleotides is well balanced.

In Figs.~\ref{fig3} to~\ref{fig5}, the dots depict the results of numerical simulation with Gillespie's algorithm and the lines the predictions of the theoretical method of Section~\ref{sec:theory}, confirming the excellent agreement between the two methods.  However, the theoretical method is computationallly faster than the method of numerical simulation.

\subsection{Local probabilities of copy units and mean local dwell time}

In the theory of Section~\ref{sec:theory}, the local probabilities and the mean local dwell time are given by the formulas~(\ref{mu(m;l)-scalar}) and~(\ref{tau_l-scalar}), respectively.  Here, we also consider a Bernoulli chain with equal probabilities $\nu({\rm A})=\nu({\rm T})=0.5$ for the template sequence.

\begin{figure}[h]
\includegraphics[width=9cm]{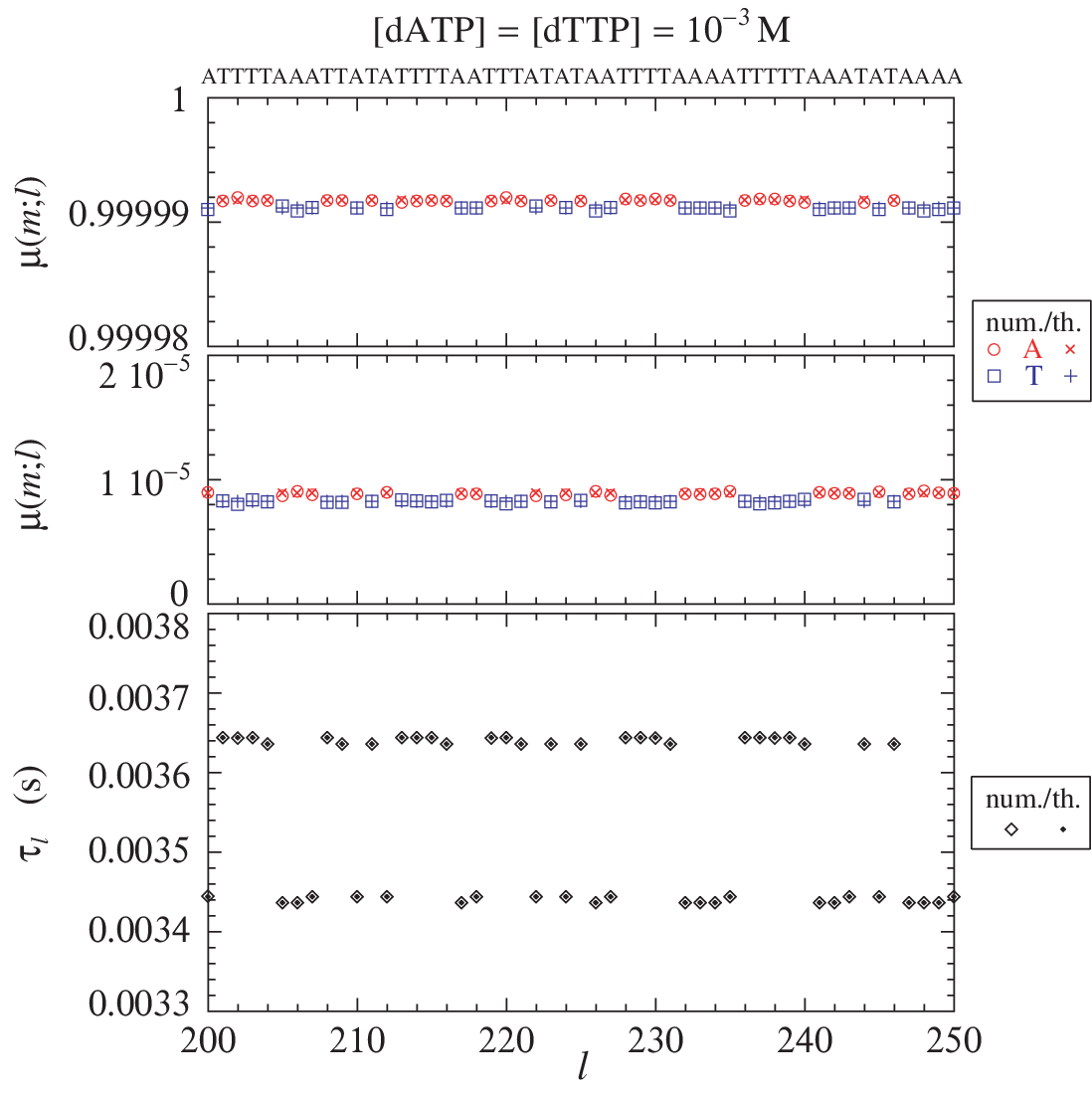}
\caption{The local probabilities $\mu(m;l)$ of the copy units $m\in\{{\rm A},{\rm T}\}$ and the mean local dwell time $\tau_l$ versus the location~$l$ along the template sequence shown above the panels for the nucleotide concentrations $[{\rm dATP}]=[{\rm dTTP}]=10^{-3}$~M.  The open symbols are the results of Gillespie's numerical algorithm using a statistics over $N_{\rm stat}=10^9$ copies.  The crosses, the pluses, and the filled diamonds are the results of theory using the formulas~(\ref{mu(m;l)-scalar}) for the local probabilities and~(\ref{tau_l-scalar}) for the mean local dwell time with $N_{\rm loop}=10$.}
\label{fig6}
\end{figure}

Figure~\ref{fig6} depicts the local probabilities of copy units $\mu({\rm A};l)$ and $\mu({\rm T};l)$, and the mean local dwell time $\tau_l$ as a function of the location~$l$ along the template sequence shown above the panels of the figure for the nucleotide concentrations $[{\rm dATP}]=[{\rm dTTP}]=10^{-3}$~M.  We observe that the local probabilities are consistent with the dominant formation of complementary base pairs A:T and T:A, as expected for a high-fidelity DNA polymerase.  In the conditions of Fig.~\ref{fig6}, the mean growth velocity and the error probability take the following values:
\bea
[{\rm dATP}]=[{\rm dTTP}]=10^{-3}\, {\rm M}: \qquad && v = 282.49 \ {\rm nt} \, {\rm s}^{-1} \, , \\
&& \eta = 8.5726 \times 10^{-6} \ {\rm nt}^{-1} \, .
\eea
In the upper panel of Fig.~\ref{fig6}, we observe that, although the concentrations are equal for the two nucleotides, the probability is slightly higher to form the correct pairs A:T than T:A.  In the middle panel, we see that, concomitantly, the incorrect pairs A:A are slightly more frequent than T:T.  Moreover, we observe in the lower panel of Fig.~\ref{fig6} that the mean local dwell time $\tau_l$ varies along the random template sequence.  These variations can be interpreted by comparing with the values for simpler template sequences.  If the template sequence is homogeneous with $n_l={\rm A}$ for all $l=1,2,3,\dots$, the mean local dwell time takes the constant value $\tau_l=0.00344$~s, which is close to the lower value in the third panel of Fig.~\ref{fig6}.  Instead, if $n_l={\rm T}$ along the whole sequence, then $\tau_l=0.00364$~s, which is close to the upper value.  If the template sequence is alternating, i.e., if $n_1n_2\cdots n_l \cdots ={\rm ATATAT}\cdots$, we have that $\tau_l=0.00344$~s for $n_l={\rm A}$ and $\tau_{l+1}=0.00364$~s for $n_{l+1}={\rm T}$, confirming the values of the homogenous sequences, which explains the structures seen in Fig.~\ref{fig6}.

\begin{figure}[h]
\includegraphics[width=9cm]{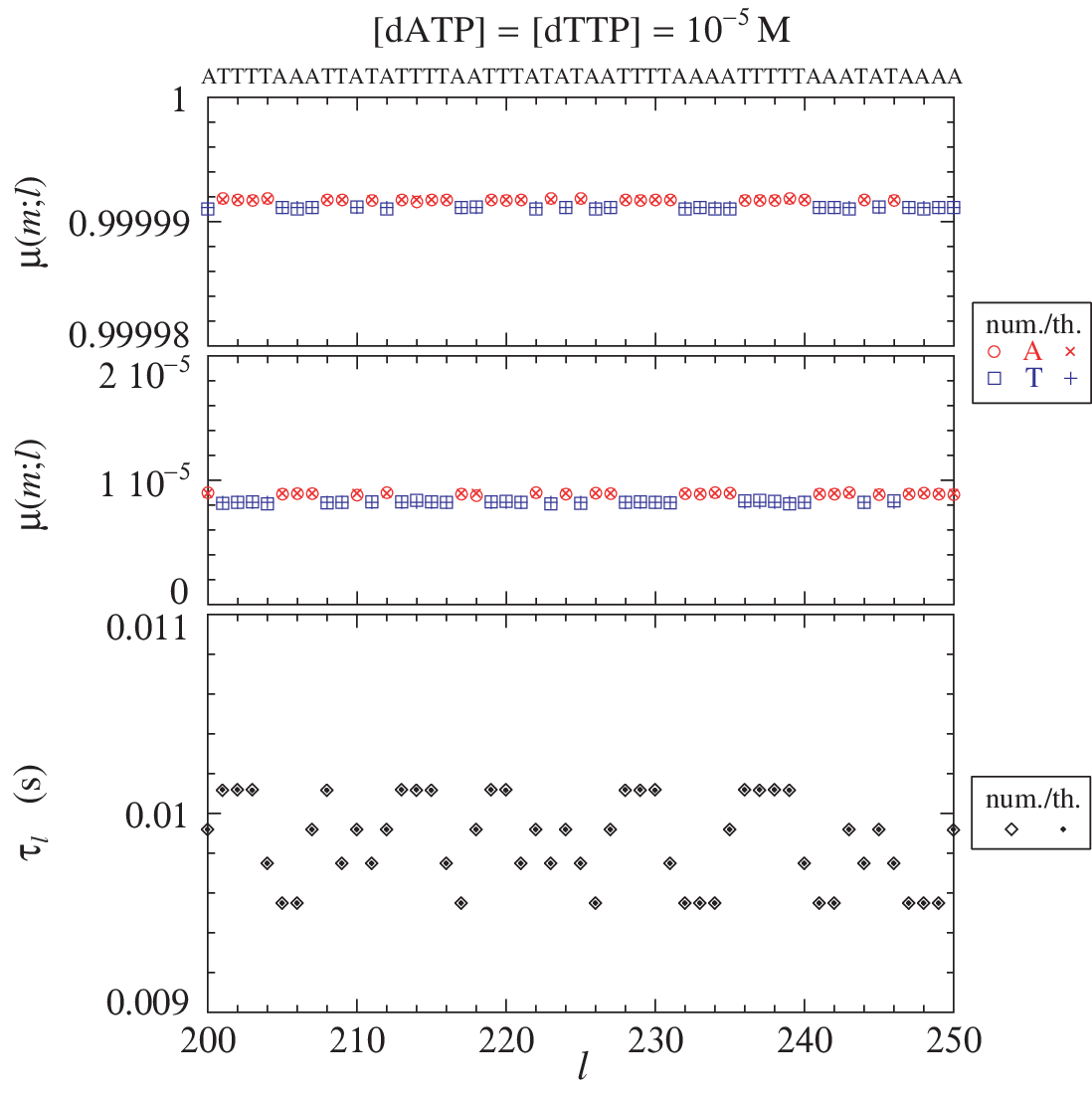}
\caption{The local probabilities $\mu(m;l)$ of the copy units $m\in\{{\rm A},{\rm T}\}$ and the mean local dwell time $\tau_l$ versus the location~$l$ along the template sequence shown above the panels for the nucleotide concentrations $[{\rm dATP}]=[{\rm dTTP}]=10^{-5}$~M.  The open symbols are the results of Gillespie's numerical algorithm using a statistics over $N_{\rm stat}=10^9$ copies.  The crosses, the pluses, and the filled diamonds are the results of theory using the formulas~(\ref{mu(m;l)-scalar}) for the local probabilities and~(\ref{tau_l-scalar}) for the mean local dwell time with $N_{\rm loop}=10$.}
\label{fig7}
\end{figure}

In Fig.~\ref{fig7}, the local probabilities of copy units and the mean local dwell time are shown along the same template sequence, but for lower nucleotide concentrations, in which case
\bea
[{\rm dATP}]=[{\rm dTTP}]=10^{-5}\, {\rm M}: \qquad && v = 101.71 \ {\rm nt} \, {\rm s}^{-1} \, , \\
&& \eta = 8.5701 \times 10^{-6} \ {\rm nt}^{-1} \, .
\eea
As in Fig.~\ref{fig6}, the upper and middle panels of Fig.~\ref{fig7} show that the pairs A:T and A:A have slightly higher probabilities than the pairs T:A and T:T, although the concentrations are equal for both nucleotides.  In the lower panel of Fig.~\ref{fig7}, we observe that the variations of the mean local dwell time are here more complicated than in Fig.~\ref{fig6}.  Again, these variations can be understood by comparing to the simpler template sequences.  We have that $\tau_l=0.00955$~s for the homogeneous sequence with $n_l={\rm A}$ and $\tau_l=0.01012$~s if $n_l={\rm T}$ for $l=1,2,3,\dots$.  Instead, for the alternating sequence ${\rm ATATAT}\cdots$, we now find that $\tau_l=0.00992$~s if $n_l={\rm A}$ and $\tau_{l+1}=0.00975$~s if $n_{l+1}={\rm T}$.  These results explain the  four bands of values taken by the mean local dwell time in Fig.~\ref{fig7}.  The upper band corresponds to the presence of successive units $n_l={\rm T}$ in the sequence, the lower band to $n_l={\rm A}$, and the two intermediate bands to switching between A and T.

\begin{figure}[h]
\includegraphics[width=9cm]{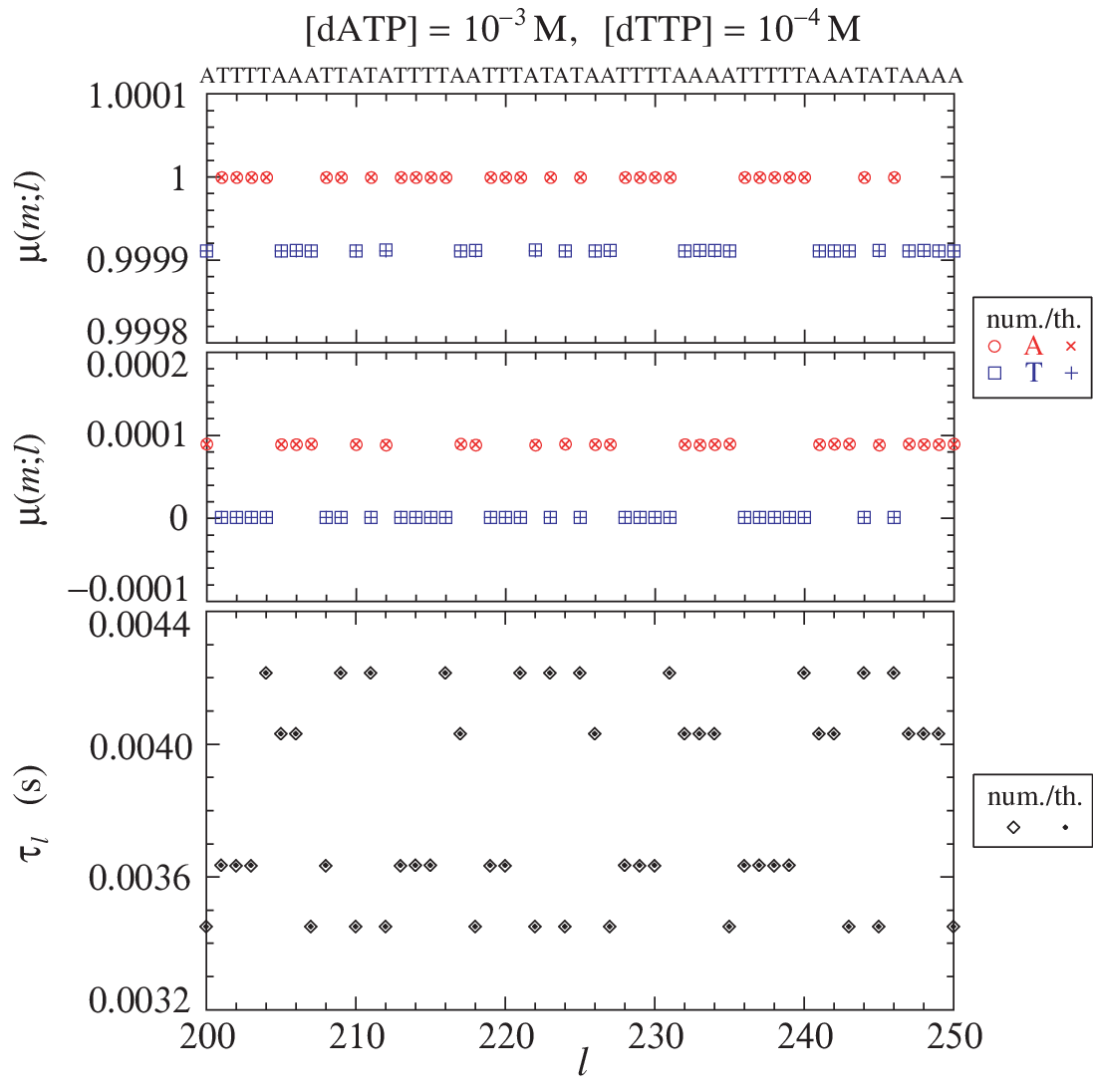}
\caption{The local probabilities $\mu(m;l)$ of the copy units $m\in\{{\rm A},{\rm T}\}$ and the mean local dwell time $\tau_l$ versus the location~$l$ along the template sequence shown above the panels for the nucleotide concentrations $[{\rm dATP}]=10^{-3}$~M and $[{\rm dTTP}]=10^{-4}$~M.  The open symbols are the results of Gillespie's numerical algorithm using a statistics over $N_{\rm stat}=10^9$ copies.  The crosses, the pluses, and the filled diamonds are the results of theory using the formulas~(\ref{mu(m;l)-scalar}) for the local probabilities and~(\ref{tau_l-scalar}) for the mean local dwell time with $N_{\rm loop}=10$.}
\label{fig8}
\end{figure}

Figure~\ref{fig8} shows the same quantities along the same random template, but for unequal concentrations such that
\bea
[{\rm dATP}]=10^{-3}\, {\rm M}\, ,\quad [{\rm dTTP}]=10^{-4}\, {\rm M}: \qquad && v = 260.95 \ {\rm nt} \, {\rm s}^{-1} \, , \\
&& \eta = 4.4945 \times 10^{-5} \ {\rm nt}^{-1} \, .
\eea
The error probability is here higher than in the two previous cases of Figs.~\ref{fig6} and~\ref{fig7}, because the nucleotide concentrations are no longer equal, which causes substitutions by the more abundant nucleotide.  In the upper and middle panels of Fig.~\ref{fig8}, we see that the pairs A:T and A:A are clearly more probable than the pairs T:A and T:T, which is explained by the larger concentration for dATP than for dTTP.  Under these conditions, we observe in the lower panel that, again, there are four bands of values for the mean local dwell time, but these bands are not in the same order as in Fig.~\ref{fig7}.  Indeed, for homogeneous sequences, we here have that $\tau_l=0.00403$~s if $n_l={\rm A}$ and $\tau_l=0.00363$~s if $n_l={\rm T}$ for $l=1,2,3,\dots$.  For the alternating sequence ${\rm ATATAT}\cdots$, we get the values $\tau_l=0.00345$~s if $n_l={\rm A}$ and $\tau_{l+1}=0.00421$~s if $n_{l+1}={\rm T}$.  Therefore, the upper and lower bands are here associated with switching and the two intermediate bands to the homogeneous sequences, which is a behavior opposite to the situation in Fig.~\ref{fig7}.

\begin{figure}[h]
\includegraphics[width=9cm]{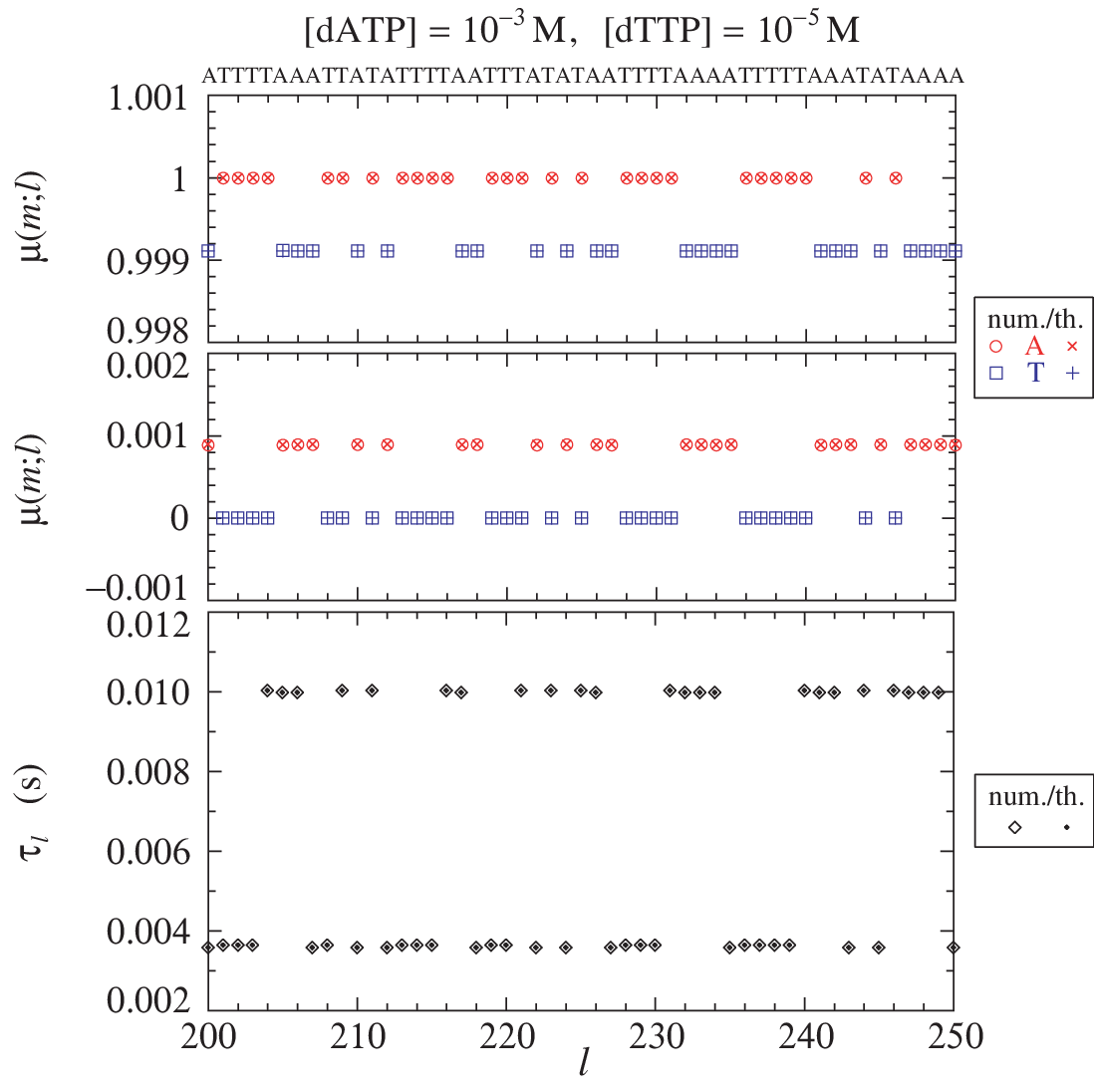}
\caption{The local probabilities $\mu(m;l)$ of the copy units $m\in\{{\rm A},{\rm T}\}$ and the mean local dwell time $\tau_l$ versus the location~$l$ along the template sequence shown above the panels for the nucleotide concentrations $[{\rm dATP}]=10^{-3}$~M and $[{\rm dTTP}]=10^{-5}$~M.  The open symbols are the results of Gillespie's numerical algorithm using a statistics over $N_{\rm stat}=10^9$ copies.  The crosses, the pluses, and the filled diamonds are the results of theory using the formulas~(\ref{mu(m;l)-scalar}) for the local probabilities and~(\ref{tau_l-scalar}) for the mean local dwell time with $N_{\rm loop}=10$.}
\label{fig9}
\end{figure}

Finally, in Fig.~\ref{fig9}, we consider the case where
\bea
[{\rm dATP}]=10^{-3}\, {\rm M}\, ,\quad [{\rm dTTP}]=10^{-5}\, {\rm M}: \qquad && v = 147.10 \ {\rm nt} \, {\rm s}^{-1} \, , \\
&& \eta = 4.4492 \times 10^{-4} \ {\rm nt}^{-1} \, .
\eea
The fidelity is here significantly lower because of the larger imbalance between the two nucleotide concentrations.
Since the concentration is larger for dATP than for dTTP, the pairs A:T and A:A are again more probable than the pairs T:A and T:T, as seen in the upper and middle panels of Fig.~\ref{fig9}.  Moreover, we recover a structure where the mean local dwell time varies between two bands of values, as depicted in the lower panel.  However, the behavior is here different from that in Fig.~\ref{fig6}.  Indeed, we now have that, for the homogeneous sequences, $\tau_l=0.00997$~s if $n_l={\rm A}$ and $\tau_l=0.00363$~s if $n_l={\rm T}$.  For the alternating sequence, the values are $\tau_l=0.00358$~s if $n_l={\rm A}$ and $\tau_{l+1}=0.01003$~s if $n_{l+1}={\rm T}$. Thus, the upper band here corresponds to values of both the homogeneous sequence with $n_l={\rm A}$ and the units $n_{l+1}={\rm T}$ of the alternating sequence.  {\it Vice versa}, the lower band corresponds to the values of both the homogeneous sequence with $n_l={\rm T}$ and the units $n_l={\rm A}$ of the alternating sequence.  In this case, switching between the two types of units gives values of the mean local dwell time that are close to those for homogeneous sequences.

The comparison demonstrates that there is excellent agreement between the results of numerical simulation with Gillespie's algorithm and those of the theoretical method of Section~\ref{sec:theory}.  We note that the theoretical method with $N_{\rm loop}=10$ is about $2\times 10^6$ faster than the method of numerical simulation with a statistics over $N_{\rm stat}=10^9$ copies to achieve the accuracy of the results shown in Figs.~\ref{fig6}-\ref{fig9}.

\section{Homogenization}
\label{sec:homog}

Homogenization is an approximation that consists in neglecting the effects of template heterogeneities during the copolymerization process catalyzed by the polymerase.  Accordingly, the template-directed copolymerization process simplifies into a free copolymerization process, in which the information about the copy growing along the template is reduced to a sequence of correct and incorrect pairs.  The correct pairs are ${\rm c}=\tilde n_l\!:\!n_l$, where $\tilde n_l$ is the nucleotide complementary to the template unit $n_l$, and the incorrect pairs are ${\rm i}=m_l\!:\!n_l$ with $m_l\ne \tilde n_l$.  Since we here consider only two types of nucleotides, the correct pairs are ${\rm c}\in\{{\rm A}\!:\!{\rm T},{\rm T}\!:\!{\rm A}\}$ and the incorrect pairs ${\rm i}\in\{{\rm A}\!:\!{\rm A},{\rm T}\!:\!{\rm T}\}$.  

\subsection{The model after homogenization}

Such an approximation is valid if several conditions are satisfied: first, the nucleotide concentrations should be equal $[{\rm dNTP}]\equiv[{\rm dATP}]=[{\rm dTTP}]$ and, secondly, the values of the rate constants should be close to each other, on the one hand, for the formation of correct pairs and, on the other hand, for that of incorrect pairs.  This is indeed the case for the values of Table~\ref{tab:pol-param}.  In this regard, we may envisage the following approximation, where the geometric mean is taken to homogenize the parameter values:
\bea
&& K_{\rm c}=K_{1,{\rm c}} = 4.03\times 10^{-4}~{\rm M} \, , \qquad\ K_{\rm i}=K_{1,{\rm i}} = 1.39\times 10^{-2}~{\rm M} \, , \label{K1c-K1i}\\
&& k^{{\rm E}\to{\rm F}}_{\rm c}=k_{2,{\rm c}} = 6500~{\rm s}^{-1} \, , \qquad\qquad k^{{\rm E}\to{\rm F}}_{\rm i}=k_{2,{\rm i}} = 170~{\rm s}^{-1} \, , \label{k2c-k2i}\\
&& k^{{\rm F}\to{\rm E}}_{\rm c}=k_{-2,{\rm c}} = 1.7~{\rm s}^{-1} \, , \qquad\qquad k^{{\rm F}\to{\rm E}}_{\rm i}=k_{-2,{\rm i}} = 340~{\rm s}^{-1} \, , \label{k-2c-k-2i}\\
&& k^{\rm p}_{+{\rm c}}=k_{3,{\rm c}} = 302~{\rm s}^{-1} \, , \qquad\qquad\quad\ k^{\rm p}_{+{\rm i}}=k_{3,{\rm i}} = 3.9~{\rm s}^{-1} \, , \label{k3c-k3i}
\eea
where the notations are respectively those of Table~\ref{tab:pol-param} and Appendix~\ref{AppB}.  Moreover, the depolymerization rate constants~(\ref{K_P-dfn}) are here given by $k^{\rm p}_{-{\rm x}} = k^{\rm p}_{+{\rm x}}/K_{\rm P}$ (for ${\rm x}={\rm c},\, {\rm i}$) with the pyrophosphorolysis constant~(\ref{K_P-value}).

We note that the Michaelis-Menten denominators~(\ref{Q-dfn}) thus take the same value given by
\be
Q=1 + \left( \frac{1}{K_{\rm c}} + \frac{1}{K_{\rm i}}\right) [{\rm dNTP}] \, ,
\label{Q-homog}
\ee
independently of the template units $n_l$ and $n_{l+1}$.  Therefore, the transition rates~(\ref{w^E>F})-(\ref{w^p-}) can here be expressed as
\be
w^{{\rm E}\to{\rm F}}_{\rm x} = k^{{\rm E}\to{\rm F}}_{\rm x} \, \frac{ [ {\rm dNTP}]}{K_{\rm x} Q } \, , \qquad
w^{{\rm F}\to{\rm E}}_{\rm x} = k^{{\rm F}\to{\rm E}}_{\rm x} \, , \qquad
w^{\rm p}_{+{\rm x}} = k^{\rm p}_{+{\rm x}} \, , \qquad
w^{\rm p}_{-{\rm x}}  = k^{\rm p}_{-{\rm x}} \, \frac{[{\rm P}]}{Q} \ , 
\qquad\mbox{for}\quad {\rm x}={\rm c},\,{\rm i} \, .
\label{w's-homog}
\ee

\subsection{The observable quantities after homogenization}

Within the approximation resulting from homogenization, the iterated function system~(\ref{IFS}) reduces to the following condition to be satisfied by the quantity $x=x_l$,
\be
\frac{\alpha_{\rm c}}{x+\beta_{\rm c}} + \frac{\alpha_{\rm i}}{x+\beta_{\rm i}} = 1 \, ,
\label{eq-x-homog}
\ee
where 
\be
\alpha_{\rm x} = \frac{w^{\rm p}_{+{\rm x}}  \, w^{{\rm E}\to{\rm F}}_{\rm x}}{w^{\rm p}_{+{\rm x}} + w^{{\rm F}\to{\rm E}}_{\rm x}} 
\qquad\mbox{and}\qquad
\beta_{\rm x} = \frac{w^{\rm p}_{-{\rm x}} \, w^{{\rm F}\to{\rm E}}_{\rm x}}{w^{\rm p}_{+{\rm x}} + w^{{\rm F}\to{\rm E}}_{\rm x}}
\qquad\mbox{for}\quad {\rm x}={\rm c},\,{\rm i} \, .
\label{alpha-beta-homog}
\ee

Accordingly, the matrices~(\ref{Y_ml-2}) simplify to
\be
\boldsymbol{\mathsf Y}_{\rm x} = 
\left[
\begin{array}{cc}
Y^{\rm E}_{\rm x} & 0 \\
Y^{\rm F}_{\rm x} & 0 \\
\end{array}
\right]
\qquad\mbox{with}\qquad
Y^{\rm E}_{\rm x} = \frac{\alpha_{\rm x}}{x+\beta_{\rm x}}
\qquad\mbox{and}\qquad
Y^{\rm F}_{\rm x} = \frac{\alpha_{\rm x}}{w^{\rm p}_{+{\rm x}}}\, \frac{x+w^{\rm p}_{-{\rm x}}}{x+\beta_{\rm x}}
\label{Y_ml-2-homog}
\ee
for ${\rm x}={\rm c},\,{\rm i}$.  Hence, the matrices~(\ref{R_l-2}) become all equal to
\be
\boldsymbol{\mathsf R}= 
\left[
\begin{array}{cc}
R^{\rm E} & 0 \\
R^{\rm F} & 0 \\
\end{array}
\right]
\qquad\mbox{with}\qquad
R^{\rm E} = 1
\qquad\mbox{and}\qquad
R^{\rm F} = \sum_{\rm x} Y^{\rm F}_{\rm x} \, .
\label{R_l-2-homog}
\ee
As a consequence, we have that $\psi^{\rm F}=R^{\rm F}\psi^{\rm E}$ and the mean local dwell time~(\ref{tau_l}) takes the uniform value $\tau=(1+R^{\rm F})/x$, so that the mean growth velocity~(\ref{v-scalar}) is now given by
\be
v = \frac{1}{\tau} = \frac{x}{1+R^{\rm F}} \, ,
\label{v-scalar-homog}
\ee
where $x$ is the positive root of Eq.~(\ref{eq-x-homog}).

\subsection{The threshold of steady growth after homogenization}

The velocity is thus equal to zero if $x=0$, which can be inserted into Eq.~(\ref{eq-x-homog}) to give the condition:
\be
\frac{\alpha_{\rm c}}{\beta_{\rm c}} + \frac{\alpha_{\rm i}}{\beta_{\rm i}} = 1 \, .
\label{v=0-homog}
\ee
Using Eqs.~(\ref{w's-homog}) and~(\ref{alpha-beta-homog}), we find the following approximation for the nucleotide concentration at the threshold of steady growth,
\be
[{\rm dNTP}]_{v=0} = \left(\frac{k^{{\rm E}\to{\rm F}}_{\rm c}}{K_{\rm c} \, k^{{\rm F}\to{\rm E}}_{\rm c}} + \frac{k^{{\rm E}\to{\rm F}}_{\rm i}}{K_{\rm i} \, k^{{\rm F}\to{\rm E}}_{\rm i}}\right)^{-1} \frac{[{\rm P}]}{K_{\rm P}} \, .
\label{dNTP-homog}
\ee
Since $K_{\rm i}=K_{1,{\rm i}}\gg K_{\rm c}=K_{1,{\rm c}}$ and $k^{{\rm E}\to{\rm F}}_{\rm c}/k^{{\rm F}\to{\rm E}}_{\rm c}=k_{2,{\rm c}}/k_{-2,{\rm c}} \gg k^{{\rm E}\to{\rm F}}_{\rm i}/k^{{\rm F}\to{\rm E}}_{\rm i}=k_{2,{\rm i}}/k_{-2,{\rm i}}$, Eq.~(\ref{dNTP-homog}) can be approximated as
\be
[{\rm dNTP}]_{v=0} \simeq \frac{K_{\rm c} \, k^{{\rm F}\to{\rm E}}_{\rm c}}{k^{{\rm E}\to{\rm F}}_{\rm c} K_{\rm P}} \, [{\rm P}] = \frac{K_{1,{\rm c}} \, k_{-2,{\rm c}}}{k_{2,{\rm c}} \, K_{\rm P}} \, [{\rm P}] = 1.8821 \times 10^{-9}~{\rm M} \, ,
\label{dNTP-homog-2}
\ee
where the numerical value is obtained with the parameter values~(\ref{K1c-K1i})-(\ref{k3c-k3i}), the pyrophosphorolysis constant~(\ref{K_P-value}), and the pyrophosphate concentration $[{\rm P}]=10^{-4}~{\rm M}$.  The approximate value~(\ref{dNTP-homog-2}) is close to the exact value~(\ref{dNTP-v=0}), showing that homogenization provides a good approximation close to the threshold of steady growth.

\subsection{The irreversible approximation after homogenization}

The irreversible approximation is here based on the assumptions that $w^{\rm p}_{-{\rm x}}=0$, so that $\beta_{\rm x}=0$ for ${\rm x}={\rm c},\,{\rm i}$.  Therefore, we find that $x=\alpha_{\rm c}+\alpha_{\rm i}$ and the mean growth velocity can thus be approximated by the following expression, which is characteristic of a Michaelis-Menten kinetics,
\be
v = \frac{k_{\rm cat}\, [{\rm dNTP}]}{[{\rm dNTP}] + K_{\rm m}}
\label{v-MM-homog}
\ee
with the parameters 
\be
k_{\rm cat} = K_{\rm m} \sum_{\rm x} \frac{k^{\rm p}_{+{\rm x}} \, k^{{\rm E}\to{\rm F}}_{\rm x}}{K_{\rm x} \left(k^{\rm p}_{+{\rm x}} + k^{{\rm F}\to{\rm E}}_{\rm x}\right)} = 287.97 \ {\rm nt} \, {\rm s}^{-1}
\label{k_cat-homog}
\ee
and
\be
K_{\rm m} = \left[ \sum_{\rm x} \frac{1}{K_{\rm x}} \left( 1 + \frac{k^{{\rm E}\to{\rm F}}_{\rm x}}{k^{\rm p}_{+{\rm x}}  + k^{{\rm F}\to{\rm E}}_{\rm x}} \right)\right]^{-1} = 1.7954 \times 10^{-5} \, {\rm M} \, .
\label{K_m-homog}
\ee
The values of these approximate parameters are also close to those obtained in Eq.~(\ref{v-MM}) without approximation.

Given that the kinetic parameters~(\ref{K1c-K1i})-(\ref{k3c-k3i}) satisfy the inequalities $K_{\rm i}=K_{1,{\rm i}}\gg K_{\rm c}=K_{1,{\rm c}}$, $k^{\rm p}_{+{\rm c}}=k_{3,{\rm c}}  \gg k^{{\rm F}\to{\rm E}}_{\rm c}=k_{-2,{\rm c}}$, and $k^{{\rm F}\to{\rm E}}_{\rm i}=k_{-2,{\rm i}} \gg k^{\rm p}_{+{\rm i}}=k_{3,{\rm i}}$, the parameters~(\ref{k_cat-homog}) and~(\ref{K_m-homog}) can be further approximated by
\be
k_{\rm cat} \simeq \frac{k^{\rm p}_{+{\rm c}} \, k^{{\rm E}\to{\rm F}}_{\rm c}}{k^{\rm p}_{+{\rm c}} + k^{{\rm E}\to{\rm F}}_{\rm c}} = \frac{k_{3,{\rm c}} \, k_{2,{\rm c}}}{k_{3,{\rm c}} + k_{2,{\rm c}}} = 288.59 \ {\rm nt} \, {\rm s}^{-1}
\label{k_cat-homog-2}
\ee
and
\be
K_{\rm m} \simeq \frac{K_{\rm c} \, k^{\rm p}_{+{\rm c}}}{k^{\rm p}_{+{\rm c}} + k^{{\rm E}\to{\rm F}}_{\rm c}} = \frac{K_{1,{\rm c}} \, k_{3,{\rm c}}}{k_{3,{\rm c}} + k_{2,{\rm c}}} = 1.7893 \times 10^{-5} \, {\rm M} \, ,
\label{K_m-homog-2}
\ee
which are still in good agreement with the previous values.

\subsection{The error probability after homogenization}

Finally, in the irreversible approximation, the error probability~(\ref{eta-scalar-0}) is given after homogenization by
\be
\eta=\frac{\alpha_{\rm i}}{\alpha_{\rm c}+\alpha_{\rm i}} \, ,
\qquad\mbox{where}\qquad
\frac{\alpha_{\rm c}}{\alpha_{\rm i}} = \frac{k^{\rm p}_{+{\rm c}}}{k^{\rm p}_{+{\rm i}}} \, \frac{k^{{\rm E}\to{\rm F}}_{\rm c}}{k^{{\rm E}\to{\rm F}}_{\rm i}} \, \frac{k^{\rm p}_{+{\rm i}} + k^{{\rm F}\to{\rm E}}_{\rm i}}{k^{\rm p}_{+{\rm c}} + k^{{\rm F}\to{\rm E}}_{\rm c}} \, \frac{K_{\rm i}}{K_{\rm c}} \, ,
\label{eta-scalar-0-homog}
\ee
so that
\be
\eta = \left(\frac{\alpha_{\rm c}}{\alpha_{\rm i}} + 1\right)^{-1} = 8.6475 \times 10^{-6} \, {\rm nt}^{-1} \, .
\label{eta-scalar-0-homog-value}
\ee
Since we have the inequalities $k^{\rm p}_{+{\rm c}}=k_{3,{\rm c}}  \gg k^{{\rm F}\to{\rm E}}_{\rm c}=k_{-2,{\rm c}}$, $k^{{\rm F}\to{\rm E}}_{\rm i}=k_{-2,{\rm i}} \gg k^{\rm p}_{+{\rm i}}=k_{3,{\rm i}}$, and $\alpha_{\rm c}\gg\alpha_{\rm i}$, we obtain the following evaluation for the error probability,
\be
\eta \simeq \frac{K_{\rm c} \, k^{{\rm E}\to{\rm F}}_{\rm i} \, k^{\rm p}_{+{\rm i}} \ \ }{K_{\rm i} \ k^{{\rm E}\to{\rm F}}_{\rm c} \, k^{{\rm F}\to{\rm E}}_{\rm i}} =\frac{K_{1,{\rm c}} \, k_{2,{\rm i}} \, k_{3,{\rm i}} \ \ }{K_{1,{\rm i}} \, k_{2,{\rm c}} \, k_{-2,{\rm i}}} = 8.6978 \times 10^{-6} \, {\rm nt}^{-1} \, .
\label{eta-scalar-homog-2}
\ee
This formula is equivalent to the inverse $\eta= D^{-1}$ of the discrimination index $D$ discussed in Refs.~\cite{J10,DKJ22}.

These results show that the theory of Section~\ref{sec:theory} is consistent with what is known in biochemistry, having the advantage of giving exact theoretical predictions also in the presence of template heterogeneities.

\section{Conclusion and perspectives}
\label{sec:conclusion}

This paper was devoted to the kinetics of DNA polymerases having several structural states, as observed for high-fidelity DNA polymerases \cite{TJ06,J10,DJ20,DJ21,DKJ22}.  To study the kinetics of such molecular machines, we used the theory developed in the companion paper for the kinetics of template-directed multistate copolymerization~\cite{paperI}, where we showed that the kinetic equations of such processes can be exactly solved in terms of matrix iterations running along the template sequence.  This theory is here applied to the kinetics of two-state DNA polymerases.  Since the reaction network~(\ref{pol-react}) of DNA replication by such polymerases is sequential, the matrix iterations reduce to scalar iterations, forming a standard iterated function system~\cite{BD85}, as for single-state DNA polymerases~\cite{G16PRL,G17PRE}.

On the one hand, this powerful theoretical method provides exact efficient formulas for the mean growth velocity of the copy, the mean local dwell time of the polymerase at every location of the template, the local probabilities of the copy units along the template sequence, and the error probability of replication by the DNA polymerase.  The exactness of these formulas is confirmed by comparison with the results of numerical simulation using Gillespie's algorithm.  In this way, we can obtain the probabilities of nucleotide substitutions with great accuracy locally along the template sequence, as demonstrated in the case of the T7 DNA polymerase.

On the other hand, the study reveals that the theoretical method based on the iterated function system is computationally much faster than the numerical simulation method and this by a factor reaching a million or more.  In this regard, the knowledge of the kinetic parameters of DNA polymerases as measured in biochemistry, combined with the knowledge of the DNA sequence of a living organism in bioinformatics, could lead to predictions on the probabilities of local nucleotide substitutions possibly causing mutations during DNA replication.  The iterative method is fast enough to compute the local substitution probabilities for large genomes.

Comparing the present results to those previously obtained for the kinetics of single-state DNA polymerases~\cite{G16PRL}, we notice significant differences in the expression of the iterated function system and in the formulas for the mean growth velocity and the local probabilities of copy units. Here, in the case of two-state kinetics, the iterated function system~(\ref{IFS}) and the local probabilities of copy units~(\ref{mu(m;l)-scalar}) are expressed in terms of the effective rates~(\ref{alpha}) and~(\ref{beta}), which combine the rates of polymerization-depolymerization with the rates of transitions between the two structural states E and F of the polymerase.  As a consequence, the formula for the mean growth velocity~(\ref{v-scalar}) differs from the formula $1/v=\langle 1/x_l\rangle$ given by Eq.~(4) of Ref.~\cite{G16PRL}, because of the extra factor $1+\psi_l^{\rm F}/\psi_l^{\rm E}=1+R_l^{\rm F}/R_l^{\rm E}$, which accounts for the time spent by the polymerase on each one of its structural states E and F.

Furthermore, we have shown that the homogenization of template heterogeneities in the DNA replication process gives relatively good approximations for the mean values defined by averaging over the template sequence, such as the nucleotide concentration at zero velocity~(\ref{dNTP-homog-2}), the mean growth velocity~(\ref{v-MM-homog}) with the parameters~(\ref{k_cat-homog-2}) and~(\ref{K_m-homog-2}), and the error probability~(\ref{eta-scalar-homog-2}).  In addition, these approximations are given by analytic formulas directly in terms of the kinetic parameters of the polymerase.  These formulas are equivalent to those that are known in biochemistry \cite{J10,DKJ22}, which supports the validity of the theory developed in the present paper.

To conclude, the iterative method we have here formulated provides a most efficient algorithm to compute the exact asymptotic solution of the kinetic equations ruling template-directed copolymerization processes by multistate as well as single-state DNA polymerases.  We leave open the issue of the possible dependence of the kinetics of multistate molecular machines on the penultimate unit.  We shall address this issue in future work.


\begin{acknowledgments}
The author thanks the Universit\'e Libre de Bruxelles (ULB) for support.
\end{acknowledgments}

\appendix

\section{Deduction of the kinetic equations}
\label{AppA}

The reaction network shown in Fig.~\ref{fig1} involves the following states:
\be
\left(m_1 \cdots m_l , \, {\rm E} \qquad\ \atop n_1\, \cdots \, n_l \, n_{l+1}\cdots\right) , \qquad
\left(m_1 \cdots m_l m_{l+1}{\rm P}, \, {\rm E} \atop n_1\, \cdots \, n_l \, n_{l+1}\cdots\ \ \right) , \qquad\mbox{and}\qquad
\left(m_1 \cdots m_l m_{l+1}{\rm P}, \, {\rm F} \atop n_1\, \cdots \, n_l \, n_{l+1}\cdots\ \ \right) , 
\label{states}
\ee
where $m_l,n_l\in\{{\rm A},{\rm T}\}$ and $l\in\{0,1,2,3,\dots\}$.  The kinetic equations ruling the time evolution of their probabilities are given by
\bea
\frac{\dd}{\dd t}\, {\cal P}_t \! \left(m_1 \cdots m_l , \, {\rm E} \qquad\ \atop n_1\, \cdots \, n_l \, n_{l+1}\cdots\right) &=& 
k^{\rm p}_{+m_l \atop \ \, n_l }
\, {\cal P}_t\!\left(m_1 \cdots m_l{\rm P} , \, {\rm F} \quad \ \ \atop n_1\, \cdots \, n_l \, n_{l+1}\cdots\right)
+\sum_{m_{l+1}} k_{-m_{l+1} \atop \ \, n_{l+1}} 
\, {\cal P}_t\!\left(m_1 \cdots m_l m_{l+1}{\rm P} , \, {\rm E} \quad \ \ \atop n_1\, \cdots \, n_l \, n_{l+1}\, n_{l+2}\cdots\right)
\nonumber\\
&& - \left( k^{\rm p}_{-m_l\atop \ \, n_l}[{\rm P}] + \sum_{m_{l+1}} k_{+m_{l+1}\atop \ \, n_{l+1}}[m_{l+1}{\rm P}] \right)  
{\cal P}_t\!\left(m_1 \cdots m_l , \, {\rm E} \qquad \ \atop n_1\, \cdots \, n_l \, n_{l+1}\cdots\right) ,
\label{kin_eq_0-E}
\eea
\bea
\frac{\dd}{\dd t}\, {\cal P}_t\!\left(m_1 \cdots m_l m_{l+1}{\rm P} , \, {\rm E} \quad\ \ \atop n_1\, \cdots \, n_l \, n_{l+1}\, n_{l+2}\cdots\right) &=& 
k_{+m_{l+1}\atop \ \, n_{l+1}}[m_{l+1}{\rm P}]
\, {\cal P}_t\!\left(m_1 \cdots m_l , \, {\rm E} \qquad\ \atop n_1\, \cdots \, n_l \, n_{l+1}\cdots\right)
+k^{{\rm F}\to{\rm E}}_{m_{l+1}\atop n_{l+1}}
\, {\cal P}_t\!\left(m_1 \cdots m_l m_{l+1}{\rm P} , \, {\rm F} \quad\ \ \atop n_1\, \cdots \, n_l \, n_{l+1}\, n_{l+2}\cdots\right)
\nonumber\\
&& - \left( k^{{\rm E}\to{\rm F}}_{m_{l+1}\atop n_{l+1}} + k_{-m_{l+1}\atop \ \, n_{l+1}}\right)  
{\cal P}_t\!\left(m_1 \cdots m_l m_{l+1}{\rm P}, \, {\rm E} \quad\ \ \atop n_1\, \cdots \, n_l \, n_{l+1}\, n_{l+2}\cdots\right) ,
\label{kin_eq_P-E}
\eea
and
\bea
\frac{\dd}{\dd t}\, {\cal P}_t\!\left(m_1 \cdots m_l m_{l+1}{\rm P} , \, {\rm F} \quad\ \ \atop n_1\, \cdots \, n_l \, n_{l+1}\, n_{l+2}\cdots\right) &=& 
k^{{\rm E}\to{\rm F}}_{m_{l+1}\atop n_{l+1}}
\, {\cal P}_t\!\left(m_1 \cdots m_l m_{l+1}{\rm P} , \, {\rm E} \quad\ \atop n_1\, \cdots \, n_l \, n_{l+1}\, n_{l+2}\cdots\right)
+k^{\rm p}_{-m_{l+1}\atop \ \, n_{l+1}}[{\rm P}]
\, {\cal P}_t\!\left(m_1 \cdots m_l m_{l+1}  , \, {\rm E} \qquad \atop n_1\, \cdots \, n_l \, n_{l+1}\, n_{l+2}\cdots\right)
\nonumber\\
&& - \left( k^{\rm p}_{+m_{l+1}\atop \ \, n_{l+1}} + k^{{\rm F}\to{\rm E}}_{m_{l+1}\atop n_{l+1}}\right)  
{\cal P}_t\!\left(m_1 \cdots m_l m_{l+1}{\rm P} , \, {\rm F} \quad\ \ \atop n_1\, \cdots \, n_l \, n_{l+1}\, n_{l+2}\cdots\right) .
\label{kin_eq_P-F}
\eea

Assuming that the Michaelis-Menten conditions~(\ref{MM-conditions}) hold, the reactions between the first and the second lines in Fig.~\ref{fig1} are fast enough to maintain a quasi-equilibrium, meaning that
\be
k_{+m_{l+1}\atop \ \, n_{l+1}}[m_{l+1}{\rm P}] \, 
{\cal P}_t\!\left(m_1 \cdots m_l , \, {\rm E} \qquad \ \atop n_1\, \cdots \, n_l \, n_{l+1}\cdots\right) \simeq
k_{-m_{l+1} \atop \ \, n_{l+1}} 
\, {\cal P}_t\!\left(m_1 \cdots m_l m_{l+1}{\rm P} , \, {\rm E} \quad \ \ \atop n_1\, \cdots \, n_l \, n_{l+1}\, n_{l+2}\cdots\right) .
\label{quasi-equil}
\ee
Furthermore, we can lump together the states of Eq.~(\ref{states}) where the polymerase is found in its structural state E and introduce
\be
 P_t(m_1 \cdots m_l ,l,{\rm E}) \equiv {\cal P}_t \! \left(m_1 \cdots m_l , \, {\rm E} \qquad\ \atop n_1\, \cdots \, n_l \, n_{l+1}\cdots\right) 
 +\sum_{m_{l+1}} {\cal P}_t\!\left(m_1 \cdots m_l m_{l+1}{\rm P} , \, {\rm E} \quad \ \ \atop n_1\, \cdots \, n_l \, n_{l+1}\, n_{l+2}\cdots\right) .
 \label{PE-dfn}
\ee
As a consequence of the quasi-equilibrium conditions~(\ref{quasi-equil}), we get
\bea
{\cal P}_t \! \left(m_1 \cdots m_l , \, {\rm E} \qquad\ \atop n_1\, \cdots \, n_l \, n_{l+1}\cdots\right) &\simeq& \frac{1}{Q_{n_{l+1}}} \, P_t(m_1 \cdots m_l ,l,{\rm E}) \, , \\
{\cal P}_t\!\left(m_1 \cdots m_l m_{l+1}{\rm P} , \, {\rm E} \quad \ \ \atop n_1\, \cdots \, n_l \, n_{l+1}\, n_{l+2}\cdots\right) &\simeq& \frac{[m_{l+1}{\rm P}]}{K_{m_{l+1}\atop n_{l+1}} \, Q_{n_{l+1}}} \, P_t(m_1 \cdots m_l ,l,{\rm E}) \, , 
\eea
in terms of the dissociation constants~(\ref{diss_csts}) and the denominators~(\ref{Q-dfn}).

Now, summing Eqs.~(\ref{kin_eq_0-E}) and~(\ref{kin_eq_P-E}) and defining
\be
 P_t(m_1 \cdots m_l ,l,{\rm F}) \equiv {\cal P}_t \! \left(m_1 \cdots m_l {\rm P} , \, {\rm F} \quad\ \atop n_1\, \cdots \, n_l \, n_{l+1}\cdots\right) ,
 \label{PF-dfn}
\ee
we obtain the kinetic equations~(\ref{kin_eq_E}) and~(\ref{kin_eq_F}) ruling the time evolution of the probabilities~(\ref{PE-dfn}) and~(\ref{PF-dfn}) with the transition rates given by Eqs.~(\ref{w^E>F})-(\ref{w^p-}) under the Michaelis-Menten conditions~(\ref{MM-conditions}).

\section{The kinetic parameters}
\label{AppB}

\subsection{Generalities}

Here, we determine the kinetic parameters of Table~\ref{tab:pol-param} for the two-state model of T7 DNA polymerase using the experimental data of Refs.~\cite{DJ20,DJ21,DKJ22}.  These data are measured with kinetic methods for a given template sequence, a given initial sequence for the copy, and a given nucleotide concentration $[{\rm dNTP}]$, which is either $[{\rm dATP}]$ or $[{\rm dTTP}]$.  These conditions allow the measurement of the kinetic parameters
\be
K_1 \equiv K_{m\atop n} \, , \qquad
k_2 \equiv k^{{\rm E}\to{\rm F}}_{m\atop n} \, , \qquad
k_{-2} \equiv k^{{\rm F}\to{\rm E}}_{m\atop n} \, , \qquad
\mbox{and}\qquad
k_3 \equiv k^{\rm p}_{+m\atop \ \, n} \, ,
\label{kin_param-dfn}
\ee
for each pair $m\!:\!n$ with $m,n\in\{{\rm A},{\rm T}\}$.  Under these specific conditions, the incorporation rate of a new copy unit is given by the usual Michaelis-Menten formula $v = k_{\rm cat}[{\rm dNTP}]/([{\rm dNTP}] + K_{\rm m})$ with
\be
k_{\rm cat} = \frac{k_2 \, k_3}{k_2+k_{-2}+k_3} 
\label{k_cat-MM}
\ee
and
\be
K_{\rm m} = K_1 \, \frac{k_{-2}+k_3}{k_2+k_{-2}+k_3} \, ,
\label{K_m-MM}
\ee
if the Michaelis-Menten conditions~(\ref{MM-conditions}) hold \cite{J92}.  Therefore, the ratio of the constants~(\ref{k_cat-MM}) and~(\ref{K_m-MM}) should be equal to
\be
\frac{k_{\rm cat}}{K_{\rm m}} = \frac{k_2 \, k_3}{K_1 \, (k_{-2}+k_3)} \, .
\label{k_cat/K_m-MM}
\ee

\subsection{The parameter values for dATP:T}

According to the data in Table~2 on p.~4 of Ref.~\cite{DKJ22}, the constants~(\ref{k_cat-MM}) and~(\ref{K_m-MM}) take the following values,
\be
k_{\rm cat} = 280 \, {\rm s}^{-1} 
\qquad\mbox{and}\qquad
K_{\rm m} = 1.8 \times 10^{-5} \, {\rm M}
\label{k_cat-K_m-dATP:T}
\ee
for this correct pair.  Furthermore, the values
\be
k_2 = 6500 \, {\rm s}^{-1} 
\qquad\mbox{and}\qquad
k_{-2} = 1.7 \, {\rm s}^{-1} 
\label{k2-k-2-dATP:T}
\ee
are reported on p.~12 of Ref.~\cite{DKJ22}.  Therefore, inverting Eq.~(\ref{k_cat-MM}), we obtain
\be
k_3 = \frac{1 + k_2^{-1} k_{-2}}{k_{\rm cat}^{-1} - k_2^{-1}} = 293 \, {\rm s}^{-1} \, .
\label{k3-dATP:T}
\ee
Next, we get from Eq.~(\ref{K_m-MM}) that
\be
K_1=  K_{\rm m} \, \frac{k_2+k_{-2}+k_3}{k_{-2}+k_3} = 4.15 \times 10^{-4} \, {\rm M} \, ,
\label{K1-dATP:T}
\ee
giving the parameter values of Table~\ref{tab:pol-param} for the pair dATP:T.

\subsection{The parameter values for dTTP:A}

For this other correct pair, we have that
\be
k_{\rm cat} = 297 \, {\rm s}^{-1} 
\qquad\mbox{and}\qquad
K_{\rm m} = 1.8 \times 10^{-5} \, {\rm M}
\label{k_cat-K_m-dTTP:A}
\ee
according to Table~2 on p.~4 of Ref.~\cite{DKJ22}.  Moreover, we assume the same values
\be
k_2 = 6500 \, {\rm s}^{-1} 
\qquad\mbox{and}\qquad
k_{-2} = 1.7 \, {\rm s}^{-1} 
\label{k2-k-2-dTTP:A}
\ee
for the two correct pairs.  In a similar way as in Eqs.~(\ref{k3-dATP:T}) and~(\ref{K1-dATP:T}), we here find that
\be
k_3 = 311 \, {\rm s}^{-1}
\qquad\mbox{and}\qquad
K_1= 3.92 \times 10^{-4} \, {\rm M} \, ,
\label{k3-K1-dATP:T}
\ee
which are given in Table~\ref{tab:pol-param} for this pair.

\subsection{The parameter values for dTTP:T}

Table~6 on p.~14 of Ref.~\cite{DKJ22} gives the following parameter values in the case of two-step nucleotide binding:
\be
K_1 = 9.3 \times 10^{-3} \, {\rm M} \, , \qquad
k_2 = 170 \, {\rm s}^{-1} \, ,
\qquad\mbox{and}\qquad
k_{-2} = 340 \, {\rm s}^{-1} \, .
\label{K1-k2-k-2-dTTP:T}
\ee
Now, taking the value $k_3 = 2.4\, {\rm s}^{-1}$ and using Eq.~(\ref{k_cat/K_m-MM}), we find that $k_{\rm cat}/K_{\rm m}=128\, {\rm M}^{-1}\, {\rm s}^{-1}$, which agrees with the experimental value $k_{\rm cat}/K_{\rm m}=130\, {\rm M}^{-1}\, {\rm s}^{-1}$ reported in Table~2 on p.~4 of Ref.~\cite{DKJ22}.  These parameter values are those given in Table~\ref{tab:pol-param} for the incorrect pair dTTP:T.

\subsection{The parameter values for dATP:A}

The lower bound on the value of $k_{\rm cat}$ given in Table~2 on p.~4 of Ref.~\cite{DKJ22} for the pair dATP:A is about $2.6$ times larger than for dTTP:T, which suggests that the rate constant $k_3$ is here larger in the same proportion.  Accordingly, we here suppose that $k_3 = 6.2\, {\rm s}^{-1}$.  In addition, we assume that the rate constants of the reaction step~2 take the same values for the two incorrect pairs, so that, here also,
\be
k_2 = 170 \, {\rm s}^{-1} \, ,
\qquad\mbox{and}\qquad
k_{-2} = 340 \, {\rm s}^{-1} \, .
\label{k2-k-2-dATP:A}
\ee
Finally, using the value $k_{\rm cat}/K_{\rm m}=147\, {\rm M}^{-1}\, {\rm s}^{-1}$ in Table~2 on p.~4 of Ref.~\cite{DKJ22} and inverting Eq.~(\ref{k_cat/K_m-MM}), we obtain
\be
K_1 = \frac{k_2 \, k_3}{K_1 \, (k_{-2}+k_3)} \left(\frac{k_{\rm cat}}{K_{\rm m}}\right)^{-1} = 2.07 \times 10^{-2} \, {\rm M} \, .
\label{K1-dATP:A}
\ee
These parameter values are reported in Table~\ref{tab:pol-param} for the incorrect pair dATP:A.

\subsection{Pyrophosphorolysis}

The experimental value~(\ref{K_P-value}) of the pyrophosphorolysis constant is given in Fig.~4 on p.~7 of Ref.~\cite{DJ21} for the correct pair dATP:T.  This value implies that the nucleotide concentration at zero velocity should be given by
\be
[{\rm dNTP}]_{v=0} \simeq \frac{K_1 \, k_{-2}}{k_2\, K_{\rm P}} \, [{\rm P}] = 1.9 \times 10^{-9}~{\rm M}
\label{dNTP-v=0-eval}
\ee
for $[{\rm P}]=10^{-4}~{\rm M}$, which is consistent with Eq.~(\ref{dNTP-homog-2}) and with previous evaluations for the T7 DNA polymerase \cite{G16a}.  Furthermore, we assume that the value~(\ref{K_P-value}) is the same for all the pairs.

\section{The regime of sublinear growth in time}
\label{App:Sublin}

As explained in the companion paper~\cite{paperI}, if the template is disordered, there exists a regime of sublinear growth in time for the mean length of the copy, going as $\langle l\rangle_t\sim t^\gamma$ with $0<\gamma <1$.  The exponent $\gamma$ can be computed as the root of the equation
\be
\left\langle\left(\frac{b_l}{a_l}\right)^\gamma \right\rangle = 1 \, ,
\label{eq-bl-al}
\ee
where the forward and backward rates are here given by
\bea
a_l &=& \frac{1}{1+R^{\rm F}_{l-1}/R^{\rm E}_{l-1}} \, \sum_{m_l} w^{{\rm E}\to{\rm F}}_{m_l\atop n_l} \, , \label{a_l}\\
b_l &=& \frac{1}{R^{\rm E}_l+R^{\rm F}_l} \, \sum_{m_l} Y^{\rm F}_{m_l,l} \, w^{{\rm F}\to{\rm E}}_{m_l\atop n_l} \, . \label{b_l}
\eea

The value $\gamma=0$ corresponds to the onset of growth, where the concentrations should satisfy the condition $\langle\ln(b_l/a_l)\rangle=0$.  Indeed, this condition is obtained by taking the limit $\gamma\to 0$ in Eq.~(\ref{eq-bl-al}), which reads $\left\langle(b_l/a_l)^\gamma \right\rangle =\left\langle\exp[\gamma\ln(b_l/a_l)] \right\rangle = 1 + \gamma \left\langle\ln(b_l/a_l)\right\rangle +O(\gamma^2) = 1$.

For the value $\gamma=1$, the concentrations should satisfy the condition $\left\langle(b_l/a_l)\right\rangle =1$, which determines the threshold of the steady-growth regime, which ends the domain of sublinear growth in time.

For the model of T7 DNA polymerase with the kinetic parameters given in Subsection~\ref{subsec:T7_DNA_pol}, the nucleotide concentrations of these two critical values are given by
\bea
&&\gamma = 0 : \qquad [{\rm dATP}]={\rm dTTP}] = 1.8838 \times 10^{-9} \, {\rm M} \, , \label{A=T,g=0}\\
&&\gamma = 1 : \qquad [{\rm dATP}]={\rm dTTP}] = 1.8845 \times 10^{-9} \, {\rm M} \, , \label{A=T,g=1}
\eea
in the same conditions as in Fig.~\ref{fig3}, where the concentrations are equal.  The numerical uncertainty on each one of these concentration values is smaller than the last digit shown.

We note that, in the approximation of homogenization of Section~\ref{sec:homog}, the rates~(\ref{a_l}) and~(\ref{b_l}) should be constant along the template sequence, i.e., $a_l=a$ and $b_l=b$ for all $l=1,2,3,\dots$. Therefore, Eq.~(\ref{eq-bl-al}) reduces to $a=b$, independently of the value of the exponent $0<\gamma <1$.  Accordingly, the regime of sublinear growth in time disappears and there is directly a transition from the stationary to the steady-growth regime at $a=b$.  Since the mean growth velocity is given by $v=a-b$ in the homogeneous case, the threshold of linear growth in time consistently coincides to $v=0$ if $a=b$ \cite{AG09}.



\end{document}